\documentclass[trackchanges]{aastex701}
\usepackage{booktabs}
\usepackage{appendix}
\usepackage{amsmath}
\usepackage{comment}
\usepackage{csquotes}
\usepackage{enumitem}
\setlist[itemize]{topsep=2pt,itemsep=2pt,parsep=0pt,partopsep=0pt}

\begin{document}

\title{CubeSats Reach the Millisecond X-Ray Domain: Crab Pulsar Timing with \emph{SpIRIT}/\emph{HERMES}}

\author[orcid=0000-0002-4773-3370]{Wladimiro Leone}
\email{wladimiro.leone@unitn.it} 
\affiliation{University of Trento, Department of Physics, Via Sommarive, 14, 38123 Povo TN, Italy}
\affiliation{University of Palermo, Dipartimento di Fisica e Chimica “E. Segrè”, Via Archirafi subcampus., Via Archirafi, 36, 90123 Palermo, Italy}
\affiliation{INAF/Osservatorio Astronomico di Trieste, via G. Tiepolo 11, I-34124 Trieste, Italy}

\author[orcid=0000-0003-1843-6139]{R. Mearns}
\email[]{}
\affiliation{School of Physics, The University of Melbourne, VIC 3010, Australia}

\author[orcid=0000-0002-3220-6375]{T. di Salvo}
\email[]{} 
\affiliation{University of Palermo, Dipartimento di Fisica e Chimica “E. Segrè”, Via Archirafi subcampus., Via Archirafi, 36, 90123 Palermo, Italy}

\author[orcid=0000-0001-5458-891X]{L. Burderi}
\email[]{} 
\affiliation{Dipartimento di Fisica, Universit\`a degli Studi di Cagliari, SP Monserrato-Sestu, km 0.7, I-09042 Monserrato,
Italy}
\affiliation{INAF/IASF Palermo, via Ugo La Malfa 153, I-90146 Palermo, Italy}

\author[orcid=0000-0001-9504-9090]{M. Thomas}
\email[]{}
\affiliation{School of Physics, The University of Melbourne, VIC 3010, Australia}

\author[orcid=0000-0001-9391-305X]{M. Trenti}
\email[]{} 
\affiliation{School of Physics, The University of Melbourne, VIC 3010, Australia}

\author[orcid=0000-0002-4031-4157]{F. Fiore}
\email[]{} 
\affiliation{INAF/Osservatorio Astronomico di Trieste, via G. Tiepolo 11, I-34124 Trieste, Italy}

\author[orcid=0000-0001-8368-8565]{E. J. Marchesini}
\email[]{} 
\affiliation{INAF/OAS, Via Piero Gobetti, 93/3, 40129 Bologna, Italy}

\author[orcid=0000-0002-4794-5453]{R. Campana}
\email[]{} 
\affiliation{INAF/OAS, Via Piero Gobetti, 93/3, 40129 Bologna, Italy}
\affiliation{INFN Sezione di Bologna, Viale Berti Pichat 6/2, 40127 Bologna, Italy}

\author[orcid=0009-0004-1606-783X]{G. Baroni}
\email[]{} 
\affiliation{INAF/Osservatorio Astronomico di Trieste, via G. Tiepolo 11, I-34124 Trieste, Italy}

\author[orcid=0009-0000-0365-5940]{M. Daf\v{c}\'{\i}kov\'{a}}
\email[]{} 
\affiliation{INAF/Osservatorio Astronomico di Trieste, via G. Tiepolo 11, I-34124 Trieste, Italy}
\affiliation{Department of Theoretical Physics and Astrophysics, Masaryk University, Kotlářská 267/2, 611 37 Brno, Czech Republic}

\author[orcid=0000-0002-2701-2998]{A. Anitra}
\email[]{} 
\affiliation{University of Palermo, Dipartimento di Fisica e Chimica “E. Segrè”, Via Archirafi subcampus., Via Archirafi, 36, 90123 Palermo, Italy}

\author[orcid=0000-0001-6096-6710]{Y. Evangelista}
\email[]{} 
\affiliation{INAF/IAPS, Via del Fosso del Cavaliere, 100, 00133 Roma, Italy}

\author[orcid=0000-0001-7916-1699]{A. Sanna}
\email[]{} 
\affiliation{Dipartimento di Fisica, Universit\`a degli Studi di Cagliari, SP Monserrato-Sestu, km 0.7, I-09042 Monserrato,
Italy}

\author[orcid=0000-0002-2734-7835]{S. Puccetti}
\email[]{} 
\affiliation{ASI - Italian Space Agency, Via del Politecnico snc, 00133, Rome, Italy}

\author[orcid=0000-0003-2882-0927]{R. Iaria}
\email[]{} 
\affiliation{University of Palermo, Dipartimento di Fisica e Chimica “E. Segrè”, Via Archirafi subcampus., Via Archirafi, 36, 90123 Palermo, Italy}

\author[orcid=0009-0002-0438-2942]{S. Barraclough}
\email[]{}
\affiliation{School of Physics, The University of Melbourne, VIC 3010, Australia}

\author[orcid=0000-0001-8431-3765]{M. Ortiz del Castillo}
\email[]{}
\affiliation{School of Physics, The University of Melbourne, VIC 3010, Australia}

\author[]{R. Bertacin}
\email[]{}
\affiliation{ASI - Italian Space Agency, Via del Politecnico snc, 00133, Rome, Italy}

\author[orcid=0000-0001-6732-3654]{P. Bellutti}
\email[]{}
\affiliation{INAF/OAB, via E. Bianchi 46
I-23807 Merate, Italy}
\affiliation{Fondazione Bruno Kessler - Sensors and Devices, Via Sommarive, 18, 38123 Povo TN, Italy}

\author[orcid=0000-0002-7283-021X]{G. Bertuccio}
\email[]{}
\affiliation{Politecnico di Milano, Dept. Electronics, Information and Bioengineering, Via Giuseppe Ponzio, 34, 20133 Milano MI}
\affiliation{ INFN Milano, Via Giovanni Celoria, 16, 20133 Milan, Italy}

\author[orcid=0000-0002-8946-552X]{A. Chapman}
\email[]{}
\affiliation{Department of Mechanical Engineering, University of Melbourne, VIC 3010, Australia}

\author[]{G. Cabras}
\email[]{}
\affiliation{Dipartimento di Fisica, Universit\`a degli Studi di Cagliari, SP Monserrato-Sestu, km 0.7, I-09042 Monserrato,
Italy}

\author[orcid=0000-0001-9110-3192]{F. Ceraudo}
\email[]{}
\affiliation{INAF/IAPS, Via del Fosso del Cavaliere, 100, 00133 Roma, Italy}

\author[orcid=0000-0002-8059-3672]{T. Chen}
\email[]{}
\affiliation{IHEP/CAS, 19B Yuquan Road, Shijingshan District, Beijing, China, 100049}

\author[orcid=0000-0003-3322-234X]{M. Citossi}
\email[]{}
\affiliation{INAF/Osservatorio Astronomico di Trieste, via G. Tiepolo 11, I-34124 Trieste, Italy}

\author[orcid=0009-0005-6714-5161]{R. Crupi}
\email[]{}
\affiliation{Department of Mathematics, Computer Science and Physics, University of Udine, via delle Scienze, 206,  33100 Udine, Italy}

\author[]{G. Della Casa}
\email[]{}
\affiliation{INAF/IAPS, Via del Fosso del Cavaliere, 100, 00133 Roma, Italy}

\author[]{E. Demenev}
\email[]{}
\affiliation{Fondazione Bruno Kessler - Sensors and Devices, Via Sommarive, 18, 38123 Povo TN, Italy}

\author[orcid=0000-0002-5132-9821]{G. Dilillo}
\email[]{}
\affiliation{INAF, Osservatorio Astronomico di Roma, Via Frascati 33, Monte Porzio Catone, Roma, 00040, Italy}
\affiliation{ASI - Italian Space Agency, Via del Politecnico snc, 00133, Rome, Italy}

\author[orcid=0000-0002-7617-3421]{M. Feroci}
\email[]{}
\affiliation{INAF/IAPS, Via del Fosso del Cavaliere, 100, 00133 Roma, Italy}

\author[orcid=0000-0002-0296-1193]{F. Ficorella}
\email[]{}
\affiliation{Fondazione Bruno Kessler - Sensors and Devices, Via Sommarive, 18, 38123 Povo TN, Italy}

\author[orcid=0000-0001-8297-1983]{M. Fiorini}
\email[]{}
\affiliation{INAF/IASF Milano, Via Alfonso Corti 12, I-20133 Milano, Italy}

\author[]{N. Gao}
\email[]{}
\affiliation{IHEP/CAS, 19B Yuquan Road, Shijingshan District, Beijing, China, 100049}

\author[]{A. Guzman}
\email[]{}
\affiliation{Institut für Astronomie und Astrophysik, Sand 1, 72076 Tübingen, Germany}

\author[]{P. Hedderman}
\email[]{}
\affiliation{Institut für Astronomie und Astrophysik, Sand 1, 72076 Tübingen, Germany}

\author[]{A. Hudrap}
\email[]{}
\affiliation{Skylabs}

\author[]{C. Labanti}
\email[]{}
\affiliation{INAF/OAS, Via Piero Gobetti, 93/3, 40129 Bologna, Italy}

\author[]{G. La Rosa}
\email[]{}
\affiliation{INAF/IASF Palermo, via Ugo La Malfa 153, I-90146 Palermo, Italy}

\author[orcid=0000-0001-6514-9672]{P. Malcovati}
\email[]{}
\affiliation{University of Pavia, Dipartimento di Ingegneria Industriale e dell’Informazione, Via Adolfo Ferrata 5, 27100 Pavia}

\author[]{J. McRobbie}
\email[]{}
\affiliation{School of Physics, The University of Melbourne, VIC 3010, Australia}

\author[orcid=0000-0002-7882-6879]{F. Mele}
\email[]{}
\affiliation{Politecnico di Milano, Dept. Electronics, Information and Bioengineering, Via Giuseppe Ponzio, 34, 20133 Milano MI}
\affiliation{INFN Milano, Via Giovanni Celoria, 16, 20133 Milan, Italy}
\affiliation{INAF/IAPS, Via del Fosso del Cavaliere, 100, 00133 Roma, Italy}

\author[]{G. Molera Calvés}
\email[]{}
\affiliation{Department of Physics, University of Tasmania, Clark Road, Hobart TAS 7005, Australia}

\author[]{J. Morgan}
\email[]{}
\affiliation{School of Physics, The University of Melbourne, VIC 3010, Australia}

\author[orcid=0000-0001-9234-7412]{G. Morgante}
\email[]{}
\affiliation{INAF/OAS, Via Piero Gobetti, 93/3, 40129 Bologna, Italy}

\author[]{B. Negri}
\email[]{}
\affiliation{ASI - Italian Space Agency, Via del Politecnico snc, 00133, Rome, Italy}

\author[orcid=0000-0002-4166-0676]{D. Novel}
\email[]{}
\affiliation{Fondazione Bruno Kessler - Sensors and Devices, Via Sommarive, 18, 38123 Povo TN, Italy}

\author[orcid=0000-0002-4426-3844]{P. Nogara}
\email[]{}
\affiliation{INAF/IASF Palermo, via Ugo La Malfa 153, I-90146 Palermo, Italy}

\author[orcid=0000-0002-9352-2355]{A. Nuti}
\email[]{}
\affiliation{INAF/IAPS, Via del Fosso del Cavaliere, 100, 00133 Roma, Italy}

\author[]{E. O'Brien}
\email[]{}
\affiliation{School of Physics, The University of Melbourne, VIC 3010, Australia}

\author[orcid=0000-0002-7397-1946]{G. Pepponi}
\email[]{}
\affiliation{Fondazione Bruno Kessler - Sensors and Devices, Via Sommarive, 18, 38123 Povo TN, Italy}

\author[orcid=0000-0003-3613-4409]{M. Perri}
\email[]{}
\affiliation{INAF/OAR, Via Frascati 33, I-00040, Monte Porzio Catone, Italy}

\author[orcid=0000-0002-4089-9503]{A. Picciotto}
\email[]{}
\affiliation{Fondazione Bruno Kessler - Sensors and Devices, Via Sommarive, 18, 38123 Povo TN, Italy}

\author[orcid=0000-0002-4222-6919]{R. Piazzolla}
\email[]{}
\affiliation{ASI - Italian Space Agency, Via del Politecnico snc, 00133, Rome, Italy}

\author[orcid=0000-0003-0377-8937]{S. Pirrotta}
\email[]{}
\affiliation{ASI - Italian Space Agency, Via del Politecnico snc, 00133, Rome, Italy}

\author[]{S. Pliego Caballero}
\email[]{}
\affiliation{Institut für Astronomie und Astrophysik, Sand 1, 72076 Tübingen, Germany}

\author[orcid=0000-0002-2723-6297]{A. Rachevski}
\email[]{}
\affiliation{INFN Trieste, Padriciano, 99, 34149 Trieste, Italy}

\author[orcid=0000-0002-7625-9903]{I. Rashevskaya}
\email[]{}
\affiliation{INFN Trieste, Padriciano, 99, 34149 Trieste, Italy}

\author[]{A. Riggio}
\email[]{}
\affiliation{Dipartimento di Fisica, Universit\`a degli Studi di Cagliari, SP Monserrato-Sestu, km 0.7, I-09042 Monserrato,
Italy}

\author[orcid=0000-0002-6216-0240]{F. Russo}
\email[]{}
\affiliation{INAF/IASF Palermo, via Ugo La Malfa 153, I-90146 Palermo, Italy}

\author[]{A. Santangelo}
\email[]{}
\affiliation{Institut für Astronomie und Astrophysik, Sand 1, 72076 Tübingen, Germany}

\author[orcid=0000-0003-3101-3966]{G. Sottile}
\email[]{}
\affiliation{INAF/IASF Palermo, via Ugo La Malfa 153, I-90146 Palermo, Italy}

\author[orcid=0000-0002-2293-212X]{C. Tenzer}
\email[]{}
\affiliation{Institut für Astronomie und Astrophysik, Sand 1, 72076 Tübingen, Germany}

\author[orcid=0009-0008-2167-078X]{Y. Tao}
\email[]{}
\affiliation{School of Physics, The University of Melbourne, VIC 3010, Australia}

\author[]{S. Trevisan}
\email[]{}
\affiliation{University of Trento, Department of Physics, Via Sommarive, 14, 38123 Povo TN, Italy}

\author[orcid=0000-0003-3855-5856]{A. Vacchi}
\email[]{}
\affiliation{Department of Mathematics, Computer Science and Physics, University of Udine, via delle Scienze, 206,  33100 Udine, Italy}

\author[orcid=0000-0003-0238-4469]{G. Zampa}
\email[]{}
\affiliation{INFN Trieste, Padriciano, 99, 34149 Trieste, Italy}

\author[]{N. Zampa}
\email[]{}
\affiliation{Department of Mathematics, Computer Science and Physics, University of Udine, via delle Scienze, 206,  33100 Udine, Italy}
\affiliation{INFN Trieste, Padriciano, 99, 34149 Trieste, Italy}

\author[orcid=0000-0002-4771-7653]{S. Xiong}
\email[]{}
\affiliation{IHEP/CAS, 19B Yuquan Road, Shijingshan District, Beijing, China, 100049}

\author[orcid=0000-0001-7599-0174]{S. Yi}
\email[]{}
\affiliation{IHEP/CAS, 19B Yuquan Road, Shijingshan District, Beijing, China, 100049}

\author[]{A. Woods}
\email[]{}
\affiliation{School of Physics, The University of Melbourne, VIC 3010, Australia}

\author[]{S. Zhang}
\email[]{}
\affiliation{IHEP/CAS, 19B Yuquan Road, Shijingshan District, Beijing, China, 100049}

\author[orcid=0000-0002-6650-3925]{N. Zorzi}
\email[]{}
\affiliation{Fondazione Bruno Kessler - Sensors and Devices, Via Sommarive, 18, 38123 Povo TN, Italy}

\begin{abstract}

The \emph{High Energy Rapid Modular Ensemble of Satellites} (\emph{HERMES}) instrument is a compact X/$\gamma$-ray spectrometer operating on board the 6U (11 kg) \emph{SpIRIT} CubeSat. The payload is particularly well suited for the observation of cosmic transients such as Gamma-Ray Bursts and bright pulsars thanks to its unique broadband sensitivity from a few keV to a few MeV and the temporal resolution down to half a microsecond.
We report here the detection of the $\sim$33~ms Crab pulsar double-peaked pulse profile obtained by considering the canonical Crab ephemerides as provided by the Jodrell Bank
catalog. We collected approximately 5.7$\cdot$10$^4$ photons from 730~s of observations, in the 3 keV -- 2 MeV energy band, during a single operation, and achieved a 5$\sigma$ pulse profile significance in the 3--11.5 keV energy band with binning at the ms scale. 
The results demonstrate that \emph{SpIRIT/HERMES} can achieve millisecond timing accuracy at high energies and, thanks to its wide field of view and broad energy band, has the potential to contribute to GRB monitoring in the near future. Such capabilities were previously the domain of flagship observatories, underscoring the performance of the \emph{HERMES} instrument with its compact form factor.




\end{abstract}

\keywords{ \uat{High Energy astrophysics}{739} -- \uat{	
Pulsar timing method}{1305} -- \uat{Spectrometers}{1554}}

\section{Introduction}



%
High-energy astrophysics in the X-ray and $\gamma$-ray bands has traditionally been driven by large space-borne observatories, whose substantial collecting area, stable pointing, and precise timing have enabled detailed studies of compact objects and fast variability. Missions such as \textit{RXTE}, \textit{NuSTAR}, and \textit{NICER} have established the current state of the art in sensitivity and absolute timing, providing key insights into pulsars, accreting systems, and high-energy transients \citep{1999Swank,Klis_2006,Madsen_2015,Vivekanand_2021}. In recent years, advances in detector technology and onboard electronics have enabled CubeSat-class missions to access the X-ray and $\gamma$-ray sky with compact detectors and increasing scientific maturity, particularly in the context of time-domain and transient astrophysics \citep{Tamagawa_2025,grbalphasmallestastrophysicalspace}.

 The \emph{High Energy Rapid Modular Ensemble of Satellites} (\emph{HERMES}) payload is an innovative and compact ($\sim$1\,U, i.e. 10$\times$10$\times$10\,cm$^{3}$ volume, with $\approx$1.6\,kg mass) X/$\gamma$-ray spectrometer operating in orbit onboard the 6U \emph{SpIRIT} (\emph{Space Industry – Responsive – Intelligent – Thermal nanosatellite}) CubeSat \citep{trenti2024}. The payload adopts, for the first time in space, a siswich (\emph{silicon sandwich}) architecture, in which Silicon Drift Detectors (SDDs) are used to both detect (i) directly cosmic X-rays and (ii) gamma-rays via optical photons produced by gamma-rays scintillation in GAGG:Ce crystals \citep{Evangelista_2020,Evangelista2024,Evangelista_2024}.
 The provides a unique broadband sensitivity from a few keVs to a few MeVs. A second key feature of the \emph{HERMES} payload is its exquisite event timing. While the siswich configuration enables fast event-by-event readout, the sub-microsecond performance is primarily set by the back-end time-tagging chain. The resulting time resolution is $\sim0.5~\mu$s, about 5--7 times better than that of previous instruments dedicated to this task \citep{Campana_2022,Dilillo_2024}.

The instrument has been developed as the core element of the \emph{HERMES} (+ \emph{SpIRIT}) Pathfinder, an in-orbit demonstration mission consisting of a constellation of six 3U and one 6U CubeSats \citep{Fiore_2020,Fiore2025}, each equipped with an identical payload \citep{Evangelista2024}. The Pathfinder aims to detect and contribute to the localization high-energy astrophysical transients, such as $\gamma$-ray bursts (GRBs), with high timing precision, broad sky coverage, and good sensitivity in terms of detection threshold photon flux (at 1 s for long
and 64 ms for short GRBs) in the 50--300 keV band: 0.8 $\rm ph \,cm^{-2}\,s^{-1}$ for long GRBs and 5.6 $\rm ph \,cm^{-2}\,s^{-1}$ for short GRBs \citep{Ghirlanda_2024}. The mission architecture leverages a distributed and modular approach to improve responsiveness and localization accuracy through inter-satellite temporal triangulation techniques \citep{2013Hurley, fiore2021distributed, Thomas_2023, Leone_2025}.

This paper presents the first timing analysis of Crab pulsar data obtained with the \emph{HERMES} payload on board \emph{SpIRIT}. The Crab is a composite system consisting of a young pulsar (PSR~B0531+21) and a surrounding synchrotron nebula, and is one of the most widely used calibration sources in high-energy astrophysics. Both the pulsar and the nebula emit across the entire electromagnetic spectrum. Because of its high and stable flux, especially in the X-ray and $\gamma$-ray bands, the Crab is considered a ``standard candle'' for instrument verification \citep{Kirsch_2005}. Furthermore, the Crab pulsation, characterized by its well-known double-peaked profile with a period of about 33~ms, is used to evaluate the timing accuracy of many instruments \citep{Rots_2004}. In this work we present the detection of the Crab pulsar profile with \emph{HERMES}. Independently, a similar result was recently obtained by another CubeSat mission, Ninjasat, using an  X-ray gas detector sensitive in the 2--50 keV band \citep{Tamagawa_2025}.  This result was achieved with an exposure of 11\,ks and thanks to a collimator that narrows the field of view (FoV) to \(\approx 2.1^{\circ}\), thereby substantially reducing the non-pulsed contribution from the cosmic X-ray background. Both results demonstrate that a CubeSat-class, low-cost mission can deliver timing performance previously achievable only with larger, dedicated observatories, marking a step toward precision time-domain astrophysics from nanosatellites. 

This paper is organized as follows. Section~2 summarizes \emph{SpIRIT}/\emph{HERMES} characteristic and science operations. Section~3 explores the Crab Pulsar timing analysis, and the Pulse Profile determination with the \emph{SpIRIT}/\emph{HERMES} non-collimated detector using a 730\,s total exposure. Section~4 illustrates the comparison of the  \emph{SpIRIT}/\emph{HERMES} timing capabilities with the \emph{NuSTAR} and \emph{NICER} observatories and validate the detected \emph{SpIRIT}/\emph{HERMES} Pulse profile by modeling it with the \emph{NuSTAR} profile.

Finally, the Appendix demonstrates analytically and via Monte Carlo tests that the relatively short exposure time considered in this analysis is expected to lead to the significant Crab detection with \emph{SpIRIT}/\emph{HERMES} inferred from the observations.

\section{\emph{SpIRIT}/\emph{HERMES}}

The first \emph{HERMES} payload has been launched and operated in space thanks to international cooperation between Australia and Italy. In this section we present an overview of the \emph{SpIRIT} mission and of the early in-orbit operations of \emph{HERMES}. 

\subsection{The \emph{SpIRIT} satellite}

The \emph{SpIRIT} satellite is a 6U-XL nanosatellite, launched into a Sun-Synchronous orbit on 1st December 2023 \citep{trenti2024}. It has been successfully and continuously operating for over 2 years in orbit at the time of writing. Australian Space Agency funded elements include the spacecraft platform, supplied by an Australian commercial provider, an instrument control unit, i.e., the Payload Management System (PMS) (\citealt{OrtizDelCastillo2025_PMS}), a thermal control unit (TheMIS; \citealt{ortizdelcastillo2025b_themis}), an Iridium and GPS payload \citep{Mearns2024_Mercury} and a Graphic Processing Unit and multi-camera system \citep{ortizdelcastillo2024a_loris}, all developed by the University of Melbourne. The \emph{HERMES} payload and the S-band module have been supplied by the Italian Space Agency through a cooperative agreement with the University of Melbourne.

\subsection{The \emph{HERMES} payload }

Each \emph{HERMES} detector comprises 60 Cerium-doped Gadolinium-Aluminium-Gallium garnet (GAGG:Ce) inorganic scintillator crystals (with a cross-sectional area of $6.94\times12.1$ mm$^2$ and a thickness of 15~mm) coupled to 12 arrays of 2$\times$5 SDDs \citep{Evangelista2024,Campana_2022} cells, each with a 25~mm$^2$ sensitive area,
segmented into four independent quadrants of 30 SDD cells each. Thanks to their sensitivity to both X-ray and optical photons, SDDs provide a dual readout channel: they detect low-energy X-rays directly in the silicon (\emph{X-mode}) and, when coupled to a scintillator, they also detect $\gamma$-rays indirectly by collecting the optical scintillation light produced in the crystal (\emph{S-mode}). In the baseline layout, each crystal is optically coupled to two SDD cells, so that the resulting light-sharing pattern enables a reliable discrimination between X- and S-events \citep{Campana_2024}.
This so-called ``siswich'' architecture offers wide spectral sensitivity, spanning a few keV to several MeV, and exceptional time resolution, reaching sub-microsecond precision, thanks to the fast response of GAGG:Ce crystals and the fine segmentation of the SDD mosaics \citep{Campana_2024}. An on-board chip-scale atomic clock provides an extremely stable 10~MHz clock (that can be synchronized with the GPS Pulse Per Second signal) for the accurate time tagging of the events \citep{Campana_2024}. The SDD signals are readout by the Application Specific Integrated Circuit (ASIC), called LYRA \citep{gandola2021multi}, a multi-chip architecture of 120 Front-End chips (LYRA-FE), placed in close proximity to the SDD anodes, and  four 32-channel Back-End chips (LYRA-BE), using an innovative current-mode transmission strategy which minimizes interchannel crosstalk while keeping a low-noise performance and a low ($<$1 mW/ch) power consumption. The payload field of view is energy-dependent: it is $\sim3$~sr in the 3--20~keV band and broadens at higher energies as the shielding becomes less effective, reaching $\sim5$~sr above $50$~keV \citep{Campana_2020,Ghirlanda_2024}.

The payload on \emph{SpIRIT} achieved first light in February 2024. Since then, it has been operated as part of the satellite commissioning phase and to achieve orbital calibration \citep{Baroni2024}. Unfortunately, prior to delivery of the payload to the University of Melbourne for integration, one of the four \emph{HERMES} quadrants was damaged, and thus has not been operational after the payload integration on the spacecraft. This results in an overall reduction of about 25\% in effective area (with an associated sensitivity reduction of $\sqrt{0.75}$), which is accounted for in the analysis presented in this paper.

During Crab pulsar observations, the nominal low-energy threshold for each active channel ranges from 3 to 7~keV, due to variations in the channel-to-channel process and the working principle of the discrimination circuit \citep{Evangelista2024}.  Accordingly, throughout this paper, we adopt 3 keV as the minimum energy of the \emph{HERMES} effective area to ensure that any potential source photons are included in the analysis.

\subsection{\emph{SpIRIT} operations}

As \emph{SpIRIT} represents the first satellite launched by the University of Melbourne and by the commercial platform provider, as well as the first flight of the \emph{HERMES} payload, a protracted commissioning period has transpired, with a number of operational limitations and anomalies accruing over the operating lifetime. These limitations and anomalies have had an impact on the execution and design of HERMES observational operations, and so they are briefly outlined here.
\begin{itemize}[topsep=2pt,itemsep=2pt,parsep=0pt,partopsep=0pt]
    
\item The spacecraft platform provides a scripting engine to organize and schedule custom operations, however these scripts are limited to a critical maximum memory size \footnote[1]{The actual value is proprietary commercial information of the platform provider and experience in orbit has shown variability depending on other processes running on board.}, which introduces significant challenges in scheduling complex operations such as those associated with a scientific instrument. 

\item The PMS instrument control unit provides an interface between all \emph{SpIRIT} payloads, including the \emph{HERMES} instrument, and the spacecraft platform, granting on-board data storage, power management, and a commanding interface via the platform scripting engine allowing flexible operations. The HERMES instrument is preferentially supplied with an uninterruptible power supply by the PMS to forestall any power supply issues \citep{OrtizDelCastillo2025_PMS}.
Unfortunately, following deployment of the second solar panel in July 2024, the PMS has been suffering from randomly occurring resets, at a rate of approximately 6 resets/hr. While a direct root cause of these resets has not been identified to date, there appears to be a strong geographic correlation, which interestingly excludes regions of high-radiation such as the South Atlantic Anomaly (Ortiz del Castillo et al., in prep). These PMS anomalies have directly affected the operational cadence of the \emph{HERMES} instrument, with early commissioning operations, such as this observation of the Crab Pulsar/Nebula, directly terminated by any single reset.

\item Since launch, \emph{SpIRIT} has been affected by a number of limitations of the Attitude and Determination Control system (ADCS). Primarily, \emph{SpIRIT} has a strong intrinsic magnetic dipole of the order of $0.6\mathrm{A m^2}$ (Wen et al. under review). When this field interacts with the Earth's field, a significant torque can be induced on the spacecraft. Consequently, the spacecraft reaction wheels must rapidly increase speed to compensate for this induced torque. Once the safety threshold of the wheels is exceeded, the spacecraft must enter a "detumbling" period in which the wheels are slowed, the spacecraft tumbles, and the angular momentum is slowly bled out of the spacecraft using magnetorquers. Due to the strength of the \emph{SpIRIT} intrinsic magnetic field, a detumbling period can occur as frequently as every orbit if \emph{SpIRIT} maintains a static attitude for an extended period. Critically, in order to achieve an acceptable level of attitude determination and control during eclipse, \emph{SpIRIT} star-tracker must be pointed as close to Zenith as possible. However, fixing both the \emph{HERMES} instrument bore-sight, and the star-tracker bore-sight (orthogonal vectors), for an extended period, results in an attitude most susceptible to the momentum build-up effect of the intrinsic dipole. This anomaly, therefore, further limits the possible operational cadence of the HERMES instrument while observing a particular astrophysical source.
\item Due to both the ADCS and PMS anomalies, the commissioning of the payload GPS timing source was not possible prior to this observation. Consequently, absolute timing information is provided to the payload by means of a \enquote{mock} GPS signal. This mock signal is synchronized to the platform's GPS clock with uncertainty of up to 1~s (the mock signal synchronization is rounded down to the nearest second) and provided to the \emph{HERMES} instrument once a second. Because following a PMS reset anomaly the \emph{HERMES} payload goes through a safe shutdown sequence, its internal atomic clock is turned off. Therefore, high timing precision for photon arrival time is limited to a single uninterrupted operation with \emph{HERMES} (i.e., consistent tracking/phase connection between observing segments in time is not ensured through the power cycling boundary).
\item \emph{SpIRIT} carries an S-Band radio for payload data downlink that was planned for use through the HERMES TP/SP ground segment. However, the link had not been commissioned at the time of these observations. Therefore, all payload data that are part of this paper were downlinked via the Telemetry \& TeleCommand link provided by the UHF radio. This limits the overall transmission rate of payload data to about 200 kB/day after accounting for higher priority spacecraft health telemetry. Consequently, the planning, execution, data retrieval, and analysis iterative loop during commissioning activities was protracted. In fact, considering the time needed for uplink of telecommands, typically one iterative set of experiments could be done on a timescale of one week.
\item Due to risk aversion to exposing the \emph{HERMES} instrument to direct Sunlight during operations in case of loss of attitude control, \emph{HERMES} operations at the time of data acquisition were limited to eclipse periods. 
\end{itemize}

We note that the presence of anomalies and in-orbit limitations compared to pre-flight nominal configuration is typical for CubeSat missions, in particular for first flights. This is exemplified by the historical failure rate of CubeSats, with about 40\% probability of infant mortality for University-led CubeSats (``dead on arrival" or loss of contact within 30 days post launch\footnote[2]{\url{https://digitalcommons.usu.edu/cgi/viewcontent.cgi?article=3415&context=smallsat}}). Therefore, by this metric, the operations for two years of a complex multi-organisation CubeSat such as \emph{SpIRIT} with both technology demonstration of payloads in orbit and scientific data acquisition --- albeit with challenges --- represent a successful outcome. 

\subsection{\emph{HERMES} observations of Crab}

\begin{figure}[h!]
    \centering
        \centering
        \includegraphics[width=0.49\linewidth]{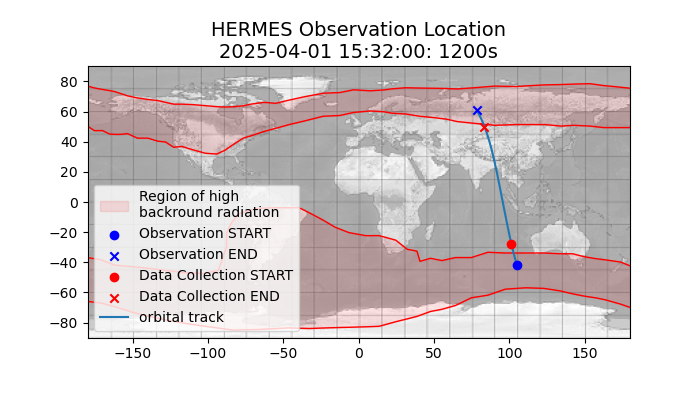}\includegraphics[width=0.49\linewidth]{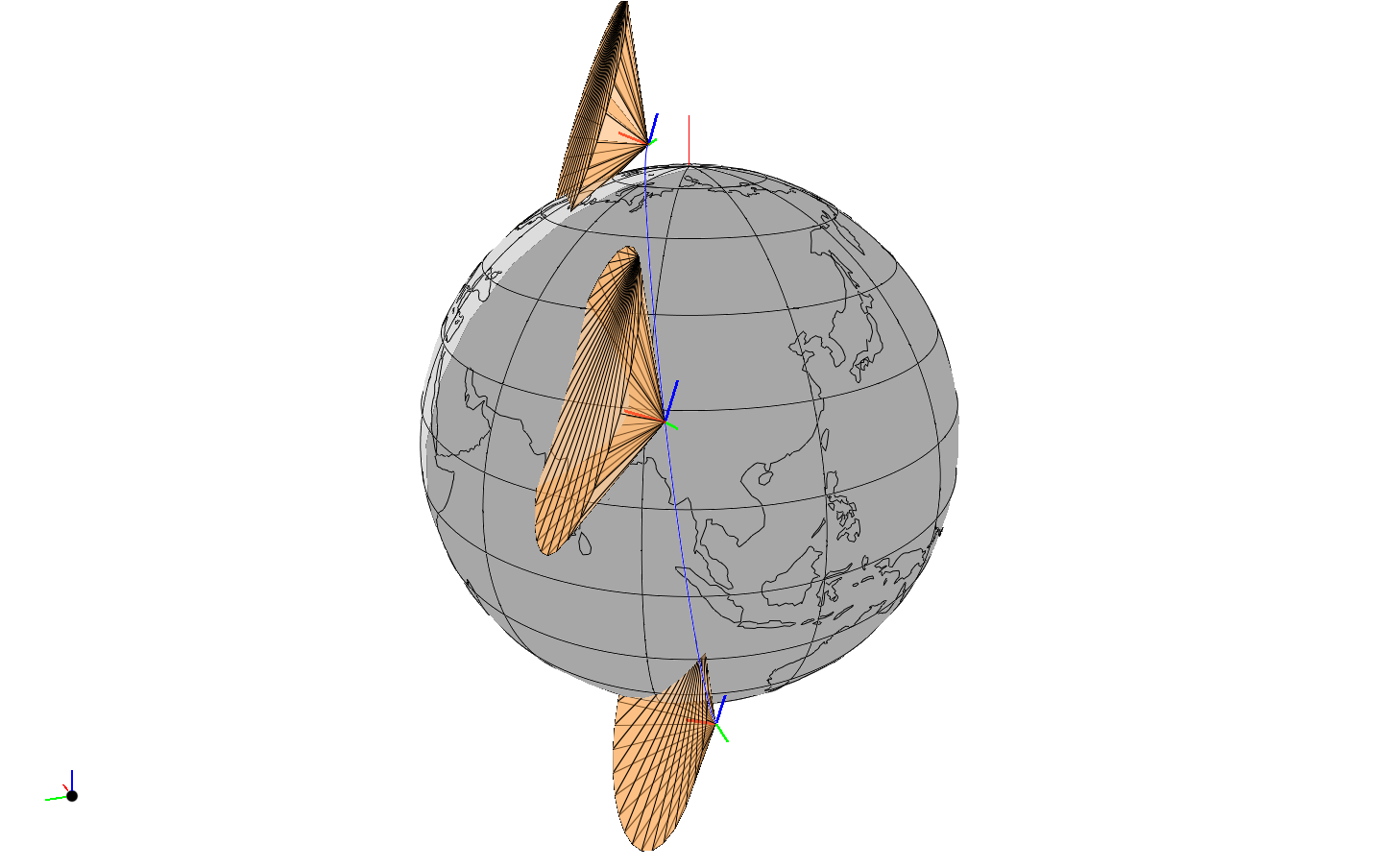} 
        \includegraphics[width=0.49\linewidth]{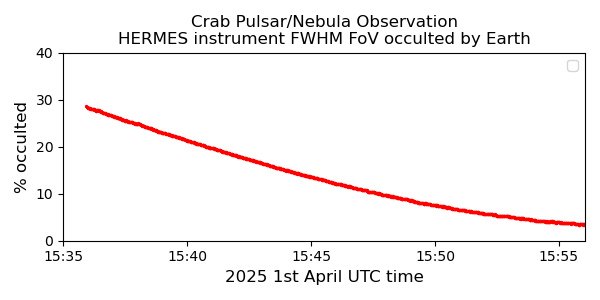}

    \caption{\emph{SpIRIT} and \emph{HERMES} instrument FoV configuration during the Crab Pulsar/Nebula observation reported here.
\textbf{Top left}: Orbital track of \emph{SpIRIT} during data collection, with high background regions shown as shaded red, data collection began at \emph{2025-04-01 15:36:06 UTC}, and ended at \emph{2025-04-01 15:56:06 UTC}.  
\textbf{Top right}: Three-dimensional display of \emph{SpIRIT} location, and \emph{HERMES} instrument FoV, during the operation.  
\textbf{Bottom}: \emph{HERMES} FoV occultation by Earth during the observation (for background noise estimation).}
    \label{fig:obs_vis}
\end{figure}

The design of the observation of the Crab Pulsar with the \emph{SpIRIT} nanosatellite is a complex multi-constraint optimisation problem; the instrument has to be operated in regions of low background radiation between the radiation belts (the contours pictured in the top left panel of Figure~\ref{fig:obs_vis}), but also in eclipse. Further, the Crab Pulsar should be located as close to the spacecraft zenith (i.e., the centre of the \emph{HERMES} field of view) as possible during the observation, to maximise the effective collecting area of photons. Competing with these constraints, however, are those to ensure the reliable operation of the platform throughout the observation. To minimise the chance of anomalous PMS resets, the observation should occur in those geographic regions with a recorded low-likelihood of resets (primarily the Indian and Atlantic oceans). Additionally, to ensure a continuous attitude solution (and thereby ensure that the Crab Pulsar remains stable in the \emph{HERMES} field of view), \emph{SpIRIT} star-tracker should point as close to Zenith as possible.

\begin{figure}[t]
    \centering
    \includegraphics[width=0.99\linewidth]{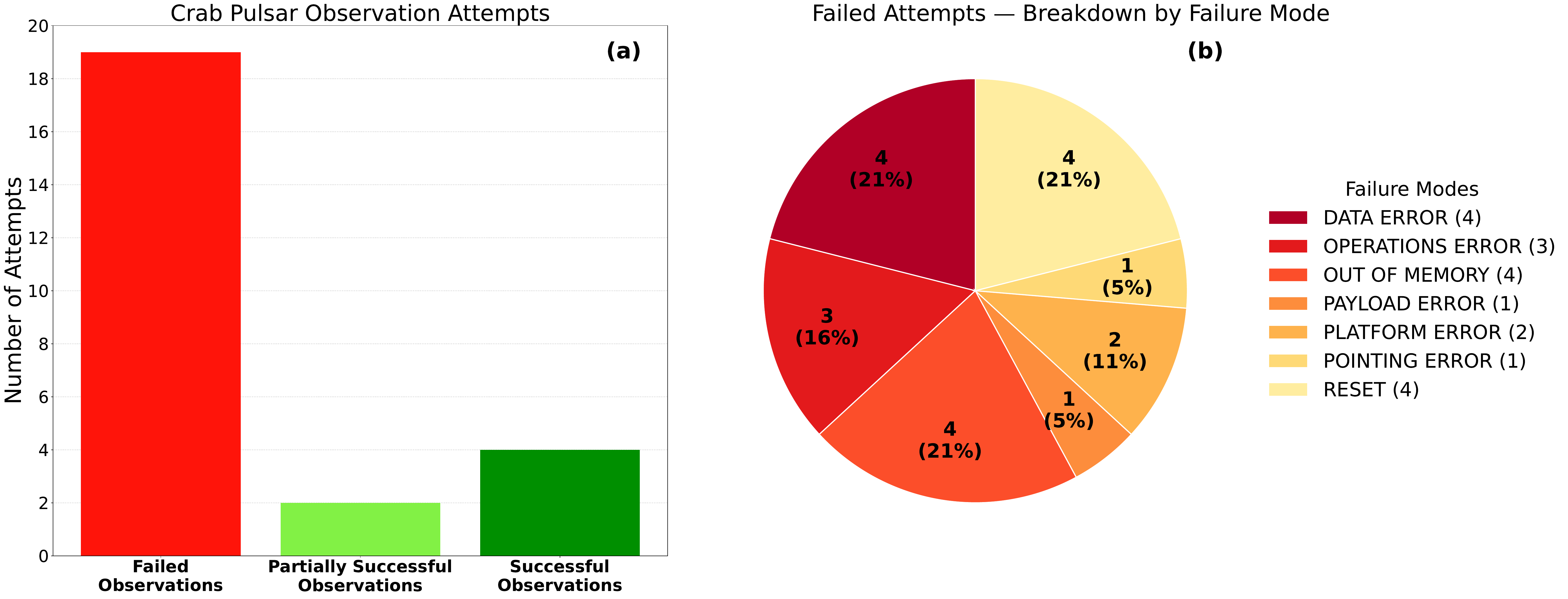}
    \caption{\textbf{(a)} Observation success statistics for the \emph{SpIRIT} Crab Pulsar/Nebula observing campaign from 2024 December 11 to 2025 April 1. \textbf{(b)} Breakdown of the failed observation attempts by failure category.}
    \label{fig:ops_success}
\end{figure}

Given these constraints the observational campaign of the Crab Pulsar/Nebula began in mid December 2024, with 25 observation attempts made over the following months to refine the process, mitigate errors within \emph{HERMES}, the payload suite, and the platform, and yield a sufficiently long dataset for the extraction of a pulsed signal. Figure~\ref{fig:ops_success} details the outcome of all observational attempts throughout this campaign. While only the primary failure mode of each observation attempt is listed, additional faults may have occurred. Of particular note as secondary failure mode is the lack of a sufficient attitude control for the duration of the observations. A brief description of each of the operation success status classifications is as follows:

\begin{itemize}
    \item \textbf{Observation Attempt Failure Mode:}
    \begin{itemize}
        \item \textbf{DATA ERROR}: Due to a design limitation of the instrument flight software in the (early) version delivered for \emph{SpIRIT}, more than 30\% of data files over a size threshold could not be offloaded from the HERMES instrument.
        \item \textbf{OPERATIONS ERROR}: Resources were not made available to the platform due to mission operations centre errors.
        \item \textbf{OUT OF MEMORY}: The operational scripts for telecommanding the observation exceeded the on-board memory limitations (this typically occurs as modifications to the operational script are rolled out to the spacecraft for the first time in order to mitigate other systematic problems).
        \item \textbf{PAYLOAD ERROR}: The \emph{HERMES} instrument failed to respond to a request within the timeout, and the observation was terminated.
        \item \textbf{PLATFORM ERROR}: The observation was not run due to an onboard corruption of the schedule file or other recoverable platform software/hardware issues.
        \item \textbf{POINTING ERROR}:  No other fault was experienced, but the spacecraft ADCS was unable to maintain an attitude lock on the Crab Pulsar/Nebula sufficient to meet the scientific observation requirements. 
        \item \textbf{RESET}:  An anomalous PMS reset was experienced, terminating the \emph{HERMES} observation.
        \end{itemize}
    \item \textbf{Partially Successful Observations}: Similar to the aforementioned DATA ERROR classification, but in this case more than 70\% of data files could be offloaded from the \emph{HERMES} instrument.
    \item \textbf{Successful Observations}:  No faults occurred during the operation, and a sufficient fraction of data files could be offloaded from the \emph{HERMES} instrument.
\end{itemize}

Most failed observations were scheduled as single $\sim$40~s exposures. Successful detections of the Crab were obtained in three separate $\sim$40~s exposures and in longer integrations of $\sim$250~s, $\sim$400~s, and $\sim$700~s, the latter being analyzed in this work. Considering the total attempted exposure time, the overall observing efficiency was approximately 50\%.

The last observation of the campaign was executed on 2025-04-01, 15:32 UTC and the instrument observed for a total of 1200~s. The orbital track of this operation is detailed in Figure~\ref{fig:obs_vis}, along with a visual representation of the HERMES instrument pointing along the orbit, and an estimate of the HERMES FoV occulted by Earth. During the observing window, the ADCS system reported that the HERMES instrument was centered within $1^\circ$ of (RA, DEC) = $(5^{h} 34.53^m, 22.01^\circ)$ (the Crab coordinates). As a caveat, we note that the accuracy and the repeatability of the ADCS has not been independently validated. Therefore, a conservative offset of $10^\circ$ from zenith has been considered on all attitude estimates. However, an offset of this magnitude has limited impact on the observations described, as the sensitivity of the instrument decreases with $\cos{\theta}$ of the offset angle for a source at zenith.

\begin{figure}[t]
    \centering
    \includegraphics[width=0.8\linewidth]{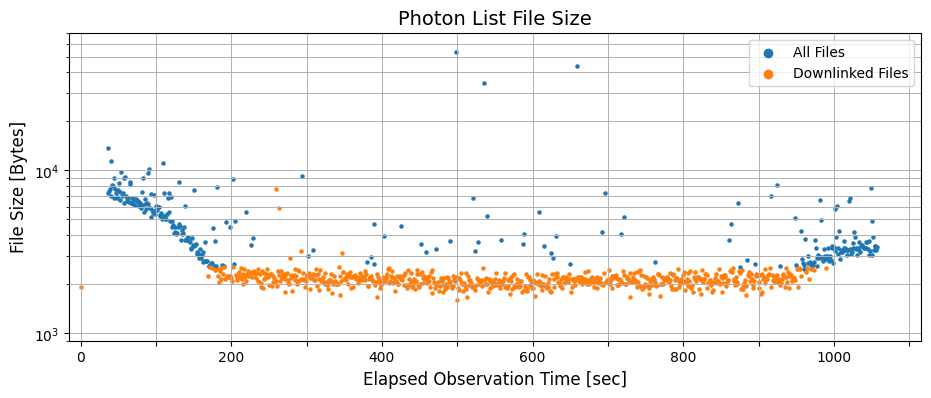}
    \caption{CALibration mode file-sizes vs. elapsed observation time. Only the files in orange were downlinked from the spacecraft, forming the non-contiguous 730~s observation dataset.}
    \label{fig:file_sizes}
\end{figure}

When correctly configured, the \emph{HERMES} instrument produces a single data file per second, timestamped according to the \emph{mock} GPS signal described above, thus the 1200~s 2025-04-01 Crab Pulsar/Nebula observation produced 1200 data files on the HERMES instrument. At the time of the observations, the on-board file processing operations were still undergoing commissioning activities. Consequently, the on-board operation to query the size of all data files required $\sim 2$ hours of computing time. Several attempts were needed to overcome the occurrence of PMS resets. Ultimately, an sufficiently complete list of 1023/1200 data files was produced. Figure~\ref{fig:file_sizes} shows output of the operation, with the size of each data file plotted against the file timestamp.

Since these data files are a contiguous list of photon energies and arrival times, the size of each file can be used as a proxy for the total number of photons collected per second, providing a pseudo-lightcurve of the observation. Noticeably, the increased count rates at the beginning and the end of the observation ($t < 170$~s, $t > 980$~s) are associated with an expanded electron belt due to the solar flare at 2025-04-01 $\sim$15:00 UTC, just prior to the observation.

Using this pseudo-lightcurve, any data files exhibiting an excessive count rate (notably while \emph{SpIRIT} was in the trailing edges of the electron belts, and additionally any temporary spikes) were excluded from the dataset. This threshold was arbitrarily set at 3 kB in size to manage UHF donwlink limitations (with a few exceptions within the first $\sim$400 seconds of the observation which were retrieved during earlier downlink attempts as part of a diagnostic process). The included/downlinked files are shown in orange in Figure~\ref{fig:file_sizes}. This process resulted in a list of 730 files, equivalent to 730 non-contiguous seconds of observational data of the Crab Pulsar/Nebula. Given that the files all originate from a single uninterrupted \emph{HERMES} operation, the internal instrument atomic clock ensures the photon events in the dataset have high-precision relative timing to the nominal sub-microsecond precision.

Datasets from observations conducted prior to 2025-04-01 have individual exposure times shorter than 400 s 
and are therefore not sufficient, when considered individually, to yield a statistically significant pulse profile, as discussed below. Moreover, these are not considered here given that the absolute timing is not available for combination across power cycling of the HERMES instrument in absence of the actual GPS signal for clock synchronization.

\section{Crab absolute timing}

The Crab pulsar Pulse Profile is folded into $n_{\mathrm{bins}} = 15$ phase channels using the effective \emph{SpIRIT} exposure of $T_\mathrm{exp} = 730\ \text{s}$ performed on 2025‑04‑01 at 15:36:06 UTC described above. Event energies are reconstructed with the standard gain/offset calibration by using the HERMESDAS reduction pipeline \citep{Puccetti2024}. We retain photons in the 3–22~keV X-mode band. We note that an X-mode detection does not automatically translate into an S-mode detection. 
For a Crab-like spectrum, the S-mode statistics in the $\sim30$~keV--2~MeV range do not allow a significant pulsed profile determination within a single $\lesssim10^3$~s exposure. 
While observation stacking would be required, the current \emph{SpIRIT/HERMES} configuration does not ensure phase connection across segments, preventing a reliable multi-segment S-mode detection.

\subsection{Barycentric Correction and Ephemeris Folding}

Pulse phases were computed by correcting the photon times of arrival with respect to the ephemeris reported in the Jodrell Bank (JB) monthly Crab database\footnote[3]{see, \url{https://www.jb.man.ac.uk/pulsar/crab/all.gro}} after extrapolating the database solution closest to the observation epoch (2025-03-17, MJD 60751). Other solutions for previous or subsequent epochs were also tested, yielding results similar to those obtained with the original solution.

The spacecraft trajectory was reconstructed by propagating the closest-in-time two‑line‑element (TLE) orbital ephemeris\footnote[4]{see, \href{https://www.space-track.org/basicspacedata/query/class/gp_history/NORAD_CAT_ID/58468/orderby/TLE_LINE1\%20ASC/EPOCH/2025-04-01--2025-04-02/format/tle}
{Space-Track TLE archive for \emph{SpIRIT} on 04-01-2025}.} set with the \textsc{SGP4} algorithm \citep{Vallado2006}.  Photon arrival times were subsequently converted to the Solar System barycenter with the \textsc{heasoft} task \texttt{barycorr}, adopting the pulsar International Celestial Reference System (ICRS) coordinates and the propagated spacecraft ephemeris. 

The Crab pulsar coordinates were taken from the Gaia DR3 catalog\footnote[5]{see, \url{https://gea.esac.esa.int/archive/}, source\_id=3403818172572314624}\citep{GaiaDR3}. The Gaia DR3 coordinates were propagated from the reference epoch J2016.0 to the observation date by applying the proper motions in right ascension and declination as:

\begin{equation}
\begin{aligned}
\alpha(t) &= \alpha_0 + \mu_{\alpha*}\,(t - t_0), \\
\delta(t) &= \delta_0 + \mu_{\delta}\,(t - t_0)
\end{aligned}
\end{equation}
where $\mu_{\alpha*}$ and $\mu_{\delta}$ are expressed in~mas\,yr$^{-1}$ and reported in the Gaia DR3 catalog.

\begin{figure}[t!]
    \centering
    \includegraphics[width=0.49\linewidth]{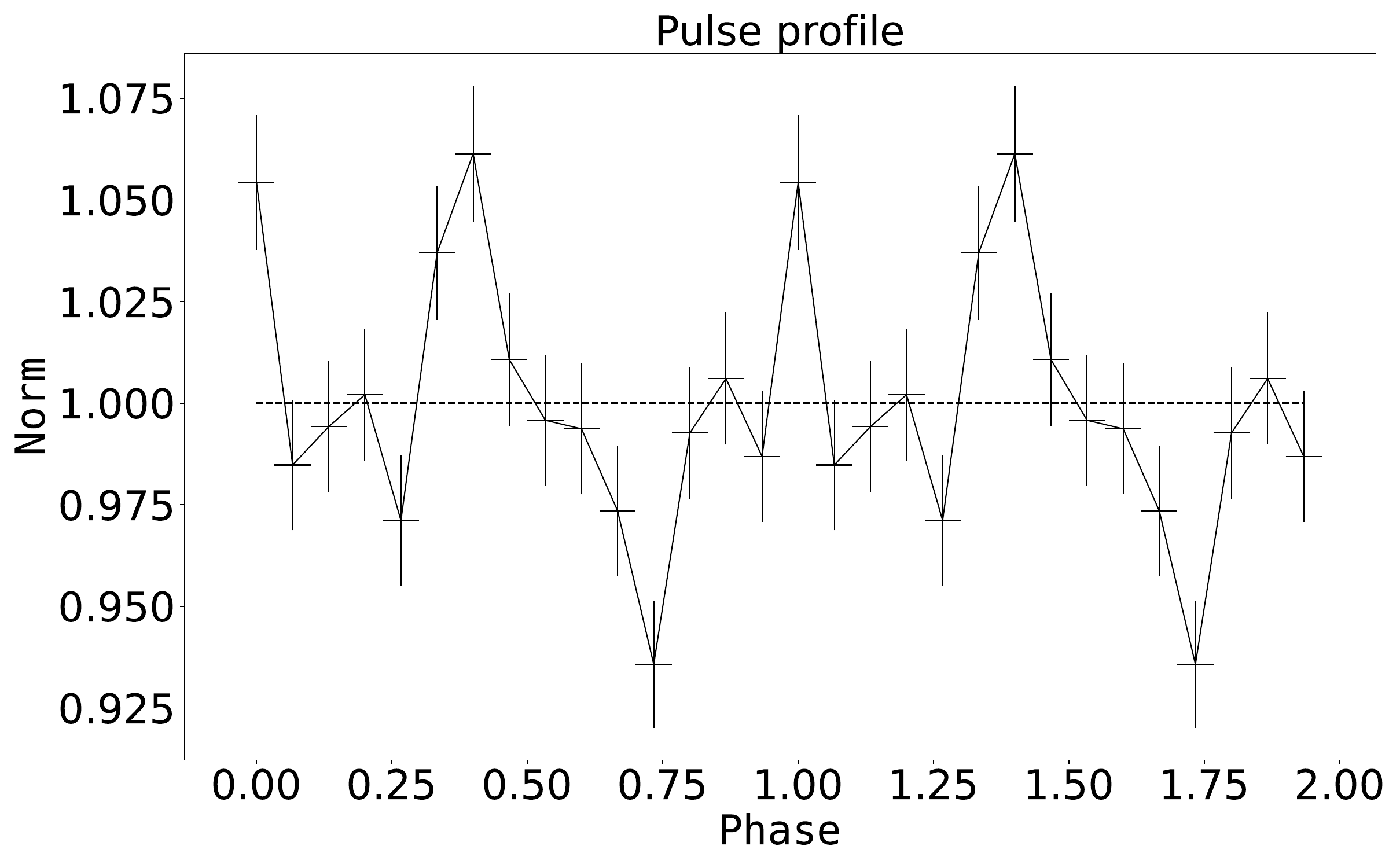}
    \includegraphics[width=0.49\linewidth]{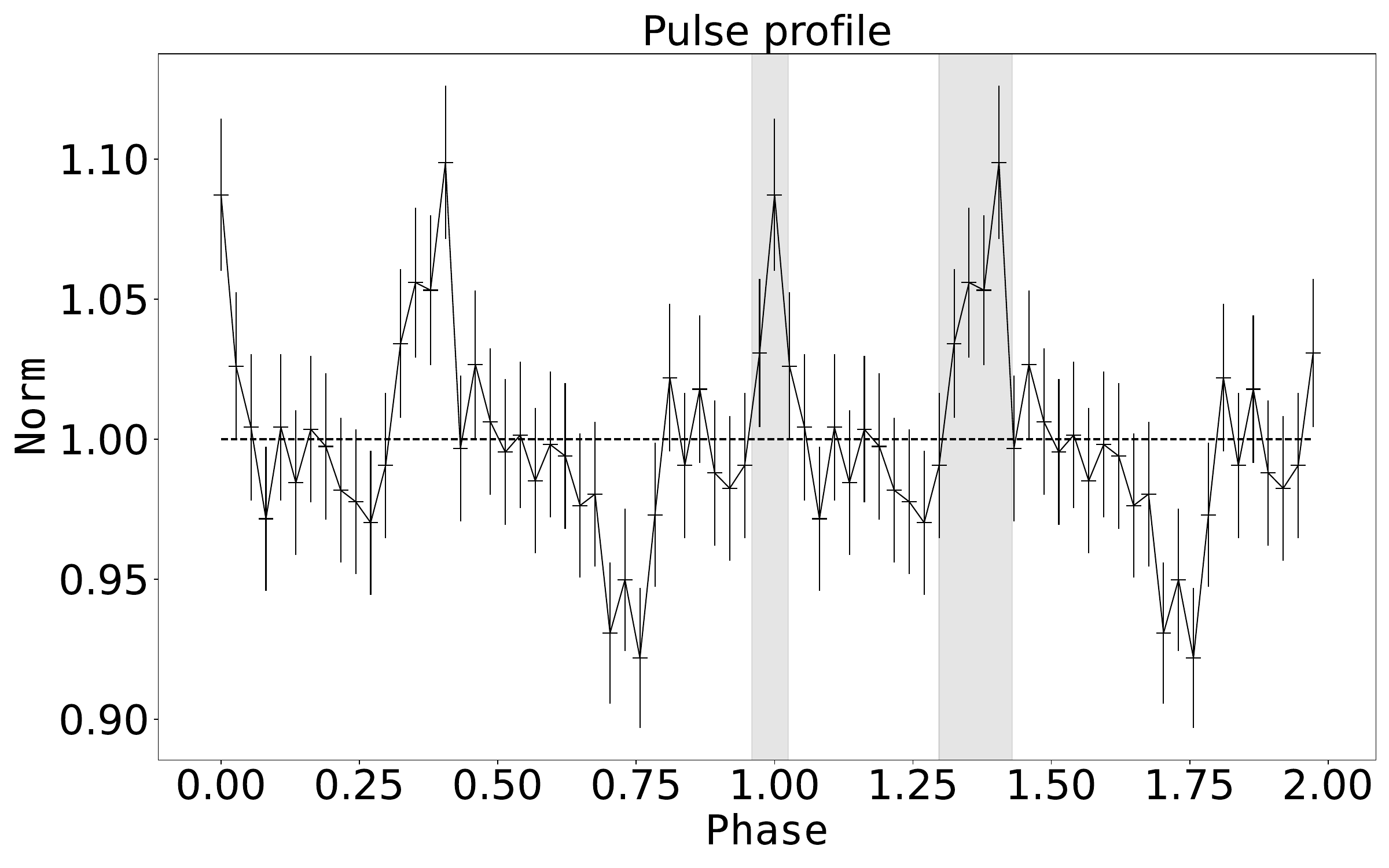}
    \caption{\emph{SpIRIT}/\emph{HERMES} CRAB pulse profiles in the 3–11.5\,keV band; barycentric-corrected times of arrival are folded with $P \simeq 33.84\,\mathrm{ms}$. \textbf{Left}: 15-channel profile with $\sim$5$\sigma$ significance. \textbf{Right}: 37-channel profile with $\sim$4$\sigma$ significance. The gray-shaded region indicates the peak locations and widths as inferred from the profile in \autoref{appendix:Pulsation significance maximization}.}
    \label{fig:CRAB_profile}
\end{figure}

Starting from the JB solution for the $P_0$ and $\dot{P}$ quoted in the catalog at $t_0=$ MJD 60751, we propagate the spin period to the epoch of the \emph{SpIRIT} observation $t$:

\begin{equation}
    P(t)\;=\;P_0\;+\;\dot{P}\,(t-t_0).
    \label{eq:P_Pdot}
\end{equation}

Hereafter, we adopt the resulting period as the HEASOFT \texttt{efold}\footnote[6]{\url{https://heasarc.gsfc.nasa.gov/docs/software/xronos/examples/efold.html}} folding period of the barycentric-corrected time-of-arrivals.


\subsection{Energy resolved profile}

We select data across several energy bands to assess the impact of noise on signal significance, while keeping the folding parameters constant. 

Consistent with the Crab pulsar spectral energy distribution peaking at 3--5\,keV \citep{Brandt2003}, we found that the pulse-profile significance is maximized when a low-energy cut is set to 3~keV. As previously discussed, although the instrument's effective area decreases between 3 and 5~keV and the nominal hardware threshold is $\sim$5~keV, channel-to-channel non-uniform response yields a spread in the \emph{effective} thresholds, leaving non-negligible acceptance down to $\sim$3~keV. 

The profile significance is evaluated by testing the null hypotheses that a flat signal is in our data (i.e., the mean of the profile) and evaluating the $\chi^2$:

\begin{equation}
\chi^{2}=\sum_{i=1}^{N}\frac{(I_{i}-\langle I\rangle)^{2}}{\sigma_{i}^{2}},
\
\end{equation}

where $I_{i}$ is the intensity of the profile and $\sigma_{i}$ the associated uncertainties. 

The 3--11.5~keV energy range was identified as the one showing the most significant pulse profile; the obtained profile with 15 folding channels is shown in \autoref{fig:CRAB_profile}, and the significance is $\sim 5\sigma$ (i.e., $\chi^2=57.5$ with dof $=14$). This is in agreement with the distribution of the expected $\chi^2$ value from the Monte Carlo simulations in \autoref{appendix:Chi_distribution}. The pulsation significance evaluated over the full X-mode band (3--20~keV) is $\sim4\sigma$ (i.e., $\chi^2=44$ for ${\rm dof}=14$). 
This reduced significance is driven by the lower Crab Pulsar statistics of the higher energy emission ($>$10 keV) and by the increasingly background-dominated count rate, with a substantial contribution from the cosmic X-ray background and from the Crab Nebula emission \citep{1999ApJ...520..124G,Kirsch_2005}.

The phase separation between the two peaks is compatible with the typical Crab profile ($\Delta\phi \approx 0.4$). In particular, the peak widths and phase positions in the 37-bin \emph{HERMES} profile (\autoref{fig:CRAB_profile}) agree with those of the schematic RXTE template shown in \autoref{appendix:Pulsation significance maximization}.

Given the uncertainty in the mock GPS time reference, the absolute phase cannot be constrained. 
Consistently, when applying several phase shift to the pulse profile in \autoref{fig:CRAB_profile}, the $\chi^{2}$ varies with only within the expected statistical scatter (i.e., $\sigma_{\chi^{2}}=\sqrt{2\,\mathrm{dof}}$). 
For completeness, the maximum $\chi^{2}$ occurs for an initial phase offset of $\tfrac{6}{15}$ of the period.

These results are nevertheless consistent with an uncertainty in the absolute phase arising from time-tagging inaccuracies, themselves a consequence of the absence of on-board GPS time discipline, given that an offset of $\pm 2.0$\,s is estimated to be present in the photon timestamps during the observation due to the initialisation of the \emph{HERMES} atomic clock with a mock GPS signal based on the main spacecraft computer clock. However, the injection of systematic shifts in the orbital file, within the range of $\pm60$\,s (i.e., corresponding to a positional uncertainty of $\pm500$\,km, which greatly exceeds the expected uncertainty), shows no impact on the pulse profile shape or significance, apart from a shift in the absolute folding phase. However, we note that while the analysis is robust against absolute shift of timestamp offsets, the absolute timing-offset uncertainties do not allow us to determine the absolute pulsar phase for the \emph{SpIRIT}  observation, as reported in the JB catalog.

 %
\subsection{Period blind search}

An independent analysis of this Crab observation was done using the \textsc{skyfield} and \textsc{astropy} libraries in Python for TLE propagation and barycentric correction. The phase $\phi$ of each photon was calculated as 
\begin{equation}
    \phi=\bmod\big(\phi_0 + \nu t + \frac{1}{2}\dot\nu t^2\big),
\end{equation}
where the frequency $\nu$ and its derivative $\dot\nu$ were taken from the JB database as previously discussed. We found the highest two-peak significance at $\phi_{0}=0.4$, consistent with the previous \texttt{barycorr}-based analysis to within half a phase-bin width $\bigl(1/(2\times 15)\approx 0.033\bigr)$. The obtained pulse profile was consistent with the one shown in \autoref{fig:CRAB_profile}.

\begin{figure}[t]
    \centering
    \includegraphics[width=0.6\linewidth]{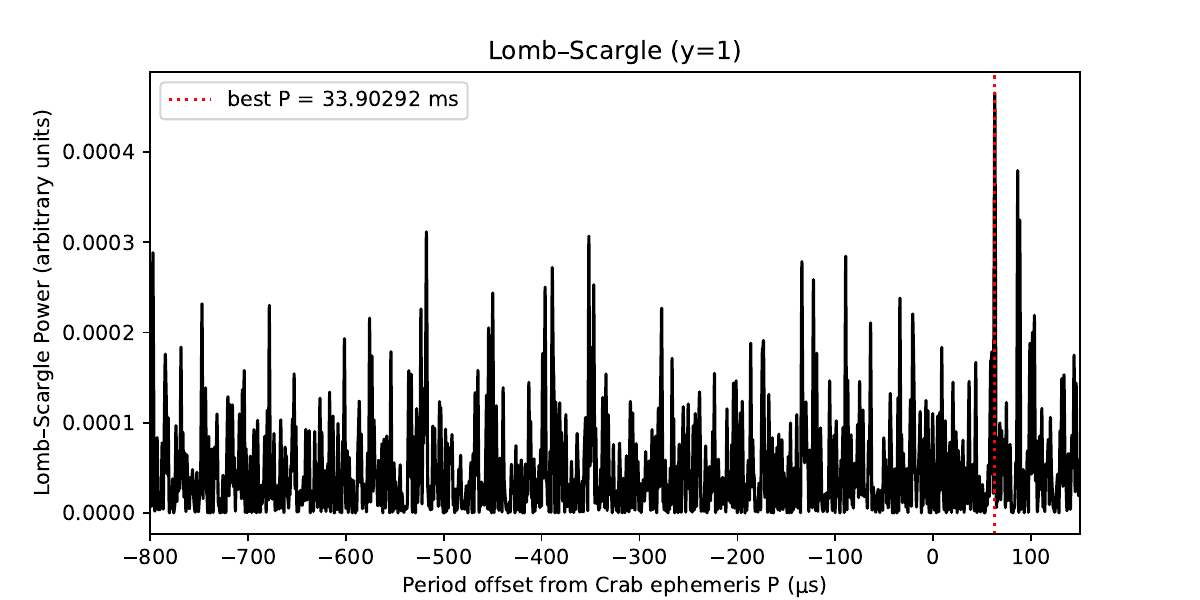}
    \caption{Blind period search with a Lomb–Scargle periodogram in the $33$–$34$~ms range: the highest peak selects the candidate period $P=33.9~\mathrm{ms}$.}
    \label{fig:Lomb–Scargle periodogram}
\end{figure}

\noindent From the barycentric-corrected photon arrival times (Skyfield--Astropy correction), we computed a Lomb--Scargle (LS) periodogram \citep{Lomb1976,Scargle1982}, assigning unit weights to each event. This unbinned approach is computationally efficient and its power normalization is independent of the source flux. We scanned all the periods in the $33$--$34$~ms window, with a 1 $\mu$s step. The statistical significance of the LS peaks was estimated via Monte Carlo simulations with $10^5$ realizations. Our analysis revealed six formally significant peaks ($>3\sigma$), with the highest one ($>3.6\sigma$) matching the Crab period, at $P_{\mathrm{LS}}=33.9\,\mathrm{ms}$. A conditioned Monte Carlo test, fixing the Crab peak and randomizing the remaining phases within the GTIs, shows that the other peaks are consistent with GTI-window sidelobes, reducing their effective significance to $<1\sigma$. The $P_{\mathrm{LS}}$ and its uncertainty were also validated via a bootstrap resampling, leading to $\sigma_{\mathrm{LS,bs}}=30\,\mu\mathrm{s}$. With this uncertainty, $P_{\mathrm{LS}}$ is consistent with the ephemeris period expected at the epoch of the \emph{SpIRIT} observation, $P(t)$, within $2\sigma_{\mathrm{LS,bs}}$.

\section{Instruments comparison}

As a further check of the derived \emph{SpIRIT}/\emph{HERMES} pulse profile, we use \emph{NICER} and \emph{NuSTAR} data to benchmark the timing performance of \emph{SpIRIT} against both instruments and to validate the recovered pulse profile by adopting the 
\emph{NICER}/\emph{NuSTAR} profiles as reference templates. The barycentric correction and folding are applied for the observation epochs as treated in the previous section.
Moreover, we used the \emph{RXTE} Crab pulse profile as a reference template to assess the significance of the pulse profile measured by \emph{SpIRIT} .

\subsection{NuSTAR and NICER data extraction}

We extract data from the closest \emph{NuSTAR} observation to the \emph{SpIRIT} Crab Pulsar observation, which started on 2025-04-05 at 22:32:21.9340 UTC with an exposure of $T_\mathrm{exp}= 14.7$~ks.
Data were retrieved from the \emph{NuSTAR}  target search website\footnote[7]{\href{https://heasarc.gsfc.nasa.gov/db-perl/W3Browse/w3table.pl?tablehead=name\%3Dnumaster&Action=More+Options}{https://heasarc.gsfc.nasa.gov/db-perl/NuSTAR\_table}}. We considered filtered data and orbital file as provided by the archive and applied the \emph{NuSTAR}  CALDB clock correction to the event list.

We also retrieved NICER data from the NICER archive\footnote[8]{\href{https://heasarc.gsfc.nasa.gov/db-perl/W3Browse/w3table.pl?tablehead=name\%3Dnicermastr&Action=More+Options}{https://heasarc.gsfc.nasa.gov/db-perl/NICER\_table}}. We consider the latest available NICER observation of the Crab pulsar, taken on 2025-02-28 18:55:18.00, with 1.4~ks exposure time, and downloaded cleaned data with the associated orbital file.

\subsection{\emph{SpIRIT} profile modeling}

We derive the \emph{NuSTAR} pulse profile $I_\mathrm{Nu}$ by considering the same number of channels as the \emph{SpIRIT} profile in \autoref{fig:CRAB_profile}. We analyze the \emph{SpIRIT} profile by fitting it with the model profile:

\begin{equation}
    I_\mathrm{mp} =A\times I_\mathrm{Nu} (x+\phi) + C,
    \label{eq:nu_model}
\end{equation}

where $A$ is the scale factor of the model and $C$ is an additive constant. The two profiles are not in phase, so we consider a free parameter $\phi$ to apply a shift to the model in order to minimize the $\chi^2$ that depends on the grade of similarity between the \emph{NuSTAR} profile and the \emph{SpIRIT} one. $I_\mathrm{mp}$ is the intensity of the model profile derived from the \emph{NuSTAR} profile $I_\mathrm{Nu}$. 
We fit the \emph{SpIRIT} profile by minimizing $\chi^{2}$ with respect to $A$, $C$, and the phase shift $\phi$, with $\phi\in[0,1)$. The best-fit phase shift is $\phi = 0.667$, while the best-fit parameters are $A = 0.41 \pm 0.073$ and $C = 0.59 \pm 0.072$. 

\begin{figure}[t]
    \centering    \includegraphics[width=0.55\linewidth]{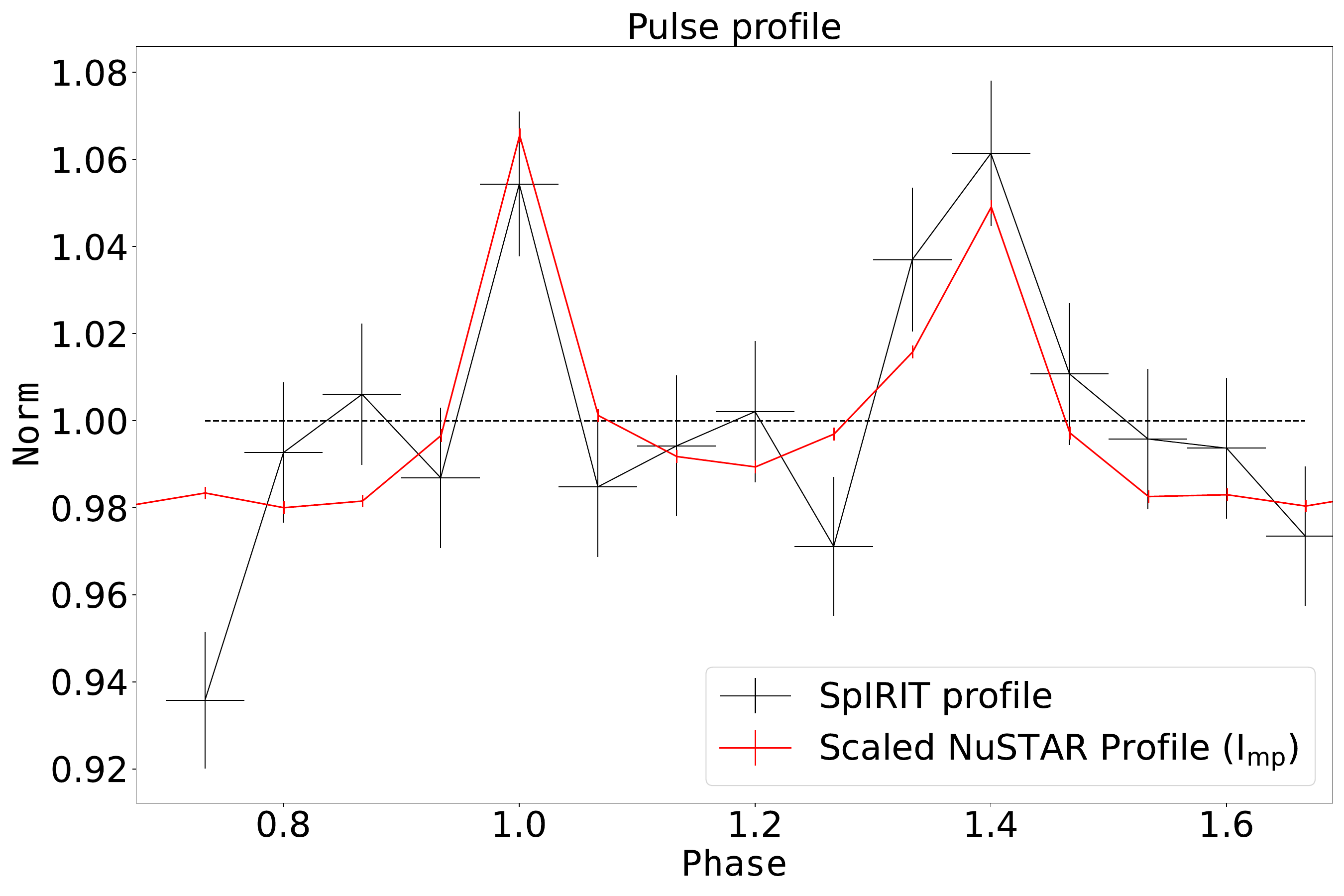}
    \caption{The Crab Pulse profile measured by \emph{SpIRIT} (15 phase bins, 3--11.5\,keV band) is fitted with the NuSTAR model in \autoref{eq:nu_model}. The solid red curve shows the NuSTAR modulated profile (15 phase bins, 3--11.5\,keV band) for the best-fit parameters A, C, and $\phi$.}
    \label{fig:profile_fit}
\end{figure}

Considering the uncertainties associated with the \emph{SpIRIT} observation and the best-fit parameters ($A, C,\phi$), \autoref{fig:profile_fit} shows the reliability of the pulse profile. 
\noindent The reduced chi-square is $\chi^2_\nu = 1.33$ with $\nu = 15-2 = 13$ degrees of freedom (i.e., $\chi^2 = 17.29$). Thus, the data are statistically consistent with the model, and there is no evidence to reject the fit at conventional significance levels.

\subsection{Epoch folding search}

Further proof of the profile accuracy is obtained by performing an epoch-folding search (\texttt{efsearch}) around the JB catalogue-expected Crab Pulsar ephemeris, as in \autoref{eq:P_Pdot}, with a 100 ns resolution. This value is the accuracy of the on-board time-stamping and therefore limits the precision of the best-period search. The \texttt{efsearch} best period is $P_\mathrm{best} = 33.8394$~ms, which is 0.5~$\mu s$ smaller with respect to the period derived from the radio ephemeris.

\begin{figure}[h!]
    \centering
        \centering
        \includegraphics[width=0.49\linewidth]{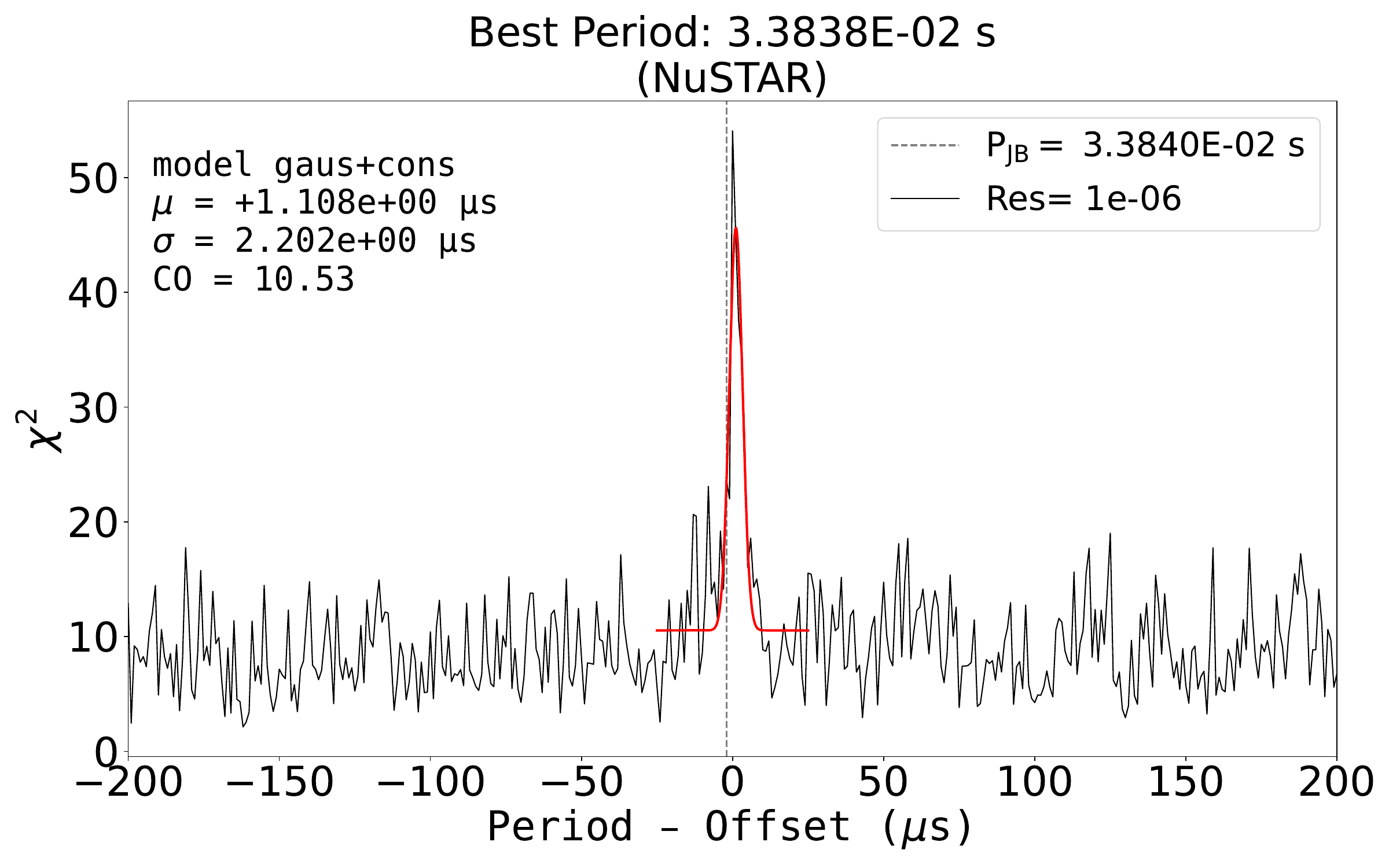}\includegraphics[width=0.49\linewidth]{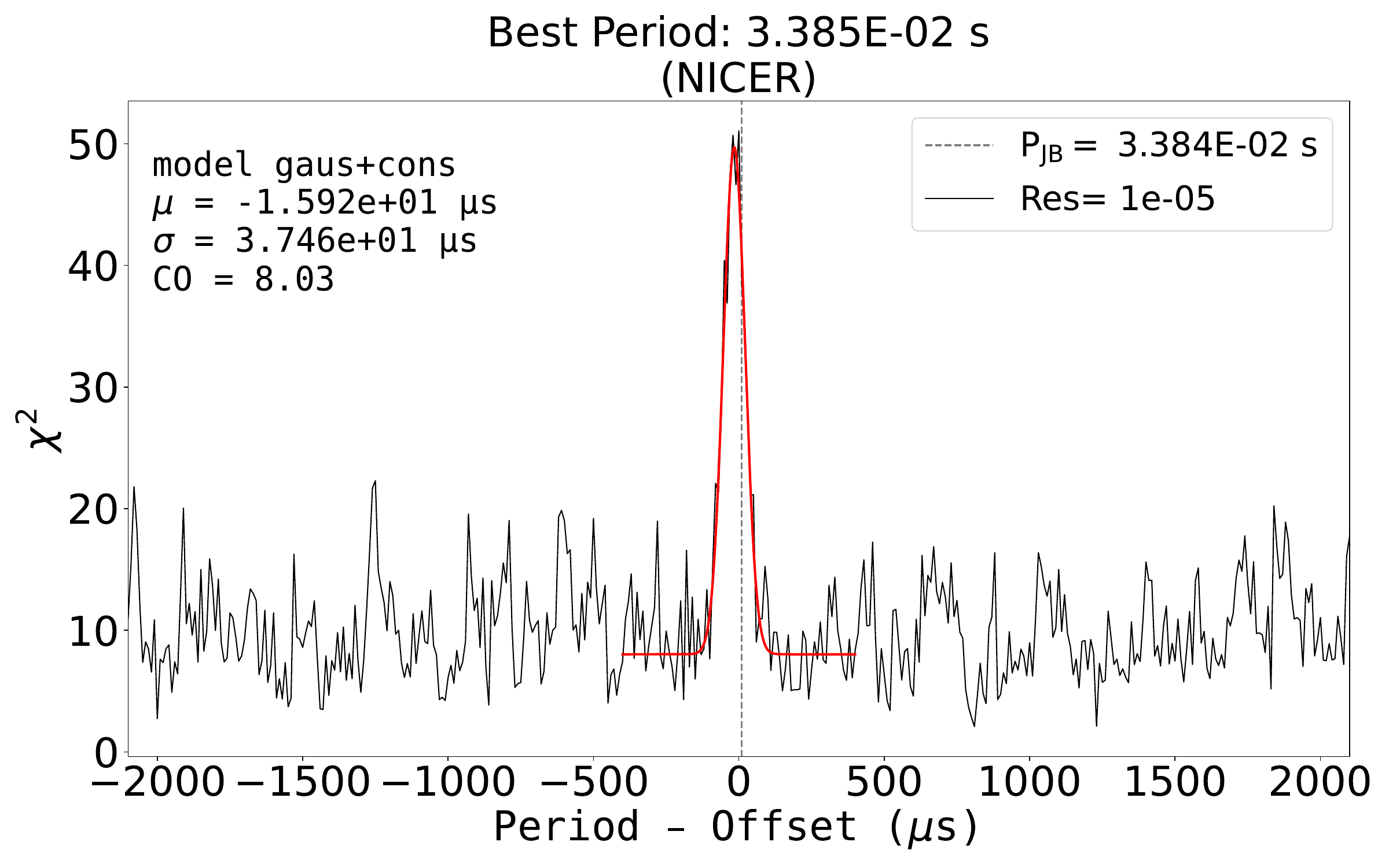} 
        \includegraphics[width=0.49\linewidth]{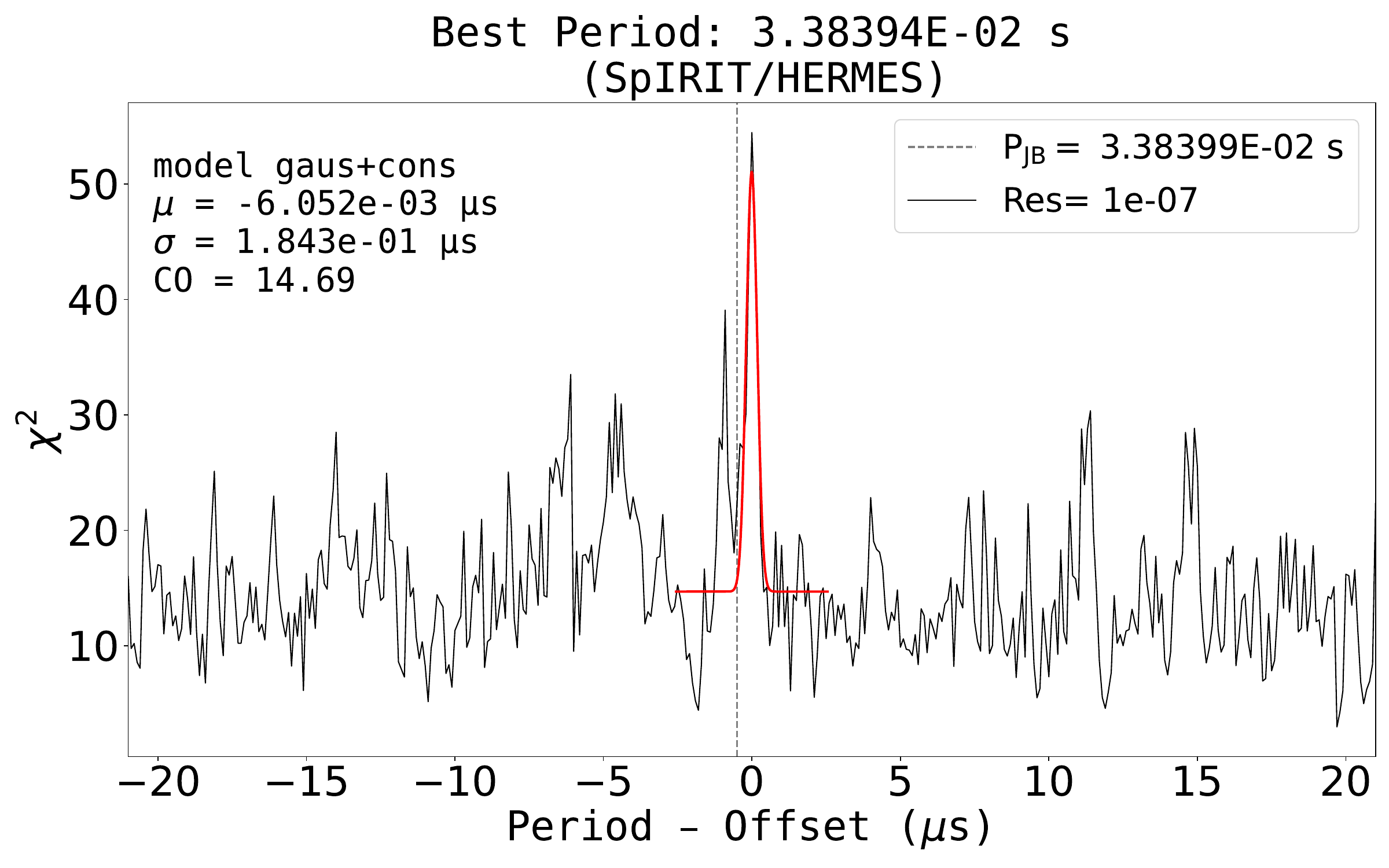}

    \caption{Epoch folding search around the expected JB catalog period at different search grid resolution, which depends on the considered exposure time. We used $n_\mathrm{bins} = 15$ for the 3--11.5 keV energy band and $P,\dot{P}$ relative to the observing epoch. Panels show the Gaussian fit of the \texttt{efsearch} for each considered instrument.}
    \label{fig:efsearch_nbin_15}
\end{figure}

The expected period is highlighted in \autoref{fig:efsearch_nbin_15}, whose difference from the best \texttt{efsearch} period is perfectly compatible with the first-order error on the period determination $\Delta P$, defined a:
\begin{equation}
    \Delta P = \frac{P^2}{T_\mathrm{exp}} \approx 1.5 \, \mu s,
\end{equation}
where $T_\mathrm{exp}$ is the observation exposure time, $T_\mathrm{exp}=730$~s.
However, the Gaussian fit to the narrow \texttt{efsearch} peak shows that the \emph{SpIRIT} best–fit period is consistent with the JB ephemeris at the observation epoch within $\sim2.5\,\sigma$, where $\sigma$ is the standard deviation of the \emph{SpIRIT/HERMES} Gaussian profile shown in \autoref{fig:efsearch_nbin_15}.

To compare the timing capabilities of \emph{SpIRIT} with \emph{NuSTAR}  and \emph{NICER} , we applied \texttt{efsearch} to \emph{NuSTAR}  and \emph{NICER}  data in the same energy band considered for the \texttt{efsearch} in \autoref{fig:efsearch_nbin_15}. We restrict the exposure so that the resulting detection significance matches that obtained with \emph{SpIRIT} (see \autoref{fig:efsearch_nbin_15}) and take into account the same Crab ephemerids used for the \emph{SpIRIT} data.

We therefore decrease the exposure time for \emph{NuSTAR}  and \emph{NICER}  data. Due to that, the \texttt{efsearch} step is:

\begin{equation}
    t_\mathrm{step} = \frac{P^2}{T_\mathrm{exp}\,n_\mathrm{bins}},
\end{equation}
which sets the period resolution shown in \autoref{fig:efsearch_nbin_15}. As expected, thanks to the larger effective area of the two instruments, they require a smaller exposure time to reach the \emph{HERMES} detection significance (\emph{NuSTAR}  $\approx 95$ s of exposure time, while for \emph{NICER}  $\approx 9$ s).

\subsection{\emph{SpIRIT} profile validation}

To validate the correct operation of the \emph{SpIRIT}/\emph{HERMES} payload and the observed pulse profile, we compare the measured significance with the predictions from the analytical and Monte Carlo analyses in \autoref{appendix:Chi_distribution}. We normalize an RXTE Crab Pulsar pulse-profile to the count rate (i.e., counts s$^{-1}$) of the \emph{HERMES} payload by considering the Cosmic X-ray Background (CXB) background and the Nebula/Pulsar contributions. First, we analytically estimate the exposure time required to guarantee a significance of at least $\gtrsim 3\sigma$ (see \autoref{appendix:Analytical}). This analysis confirms that the observed 730 s exposure is sufficient to exceed this threshold.

We then evaluate the statistical power of the detection for the fixed 730 s exposure, via Monte Carlo simulations. We generated synthetic realizations of the \emph{HERMES} pulse profile according to the Poissonian statistic, and computed the $\chi^2$ against the mean count rate for each instance. We consider as lower threshold the $\chi^2$ associated to a $3\sigma$ detection for 15 bins (i.e., $\chi^2_{\mathrm{th,3\sigma}}=35.25$). The results indicate that the probability of obtaining a $\chi^2$ value exceeding the $\chi^2_{\mathrm{th,3\sigma}}$ is $P(\chi^2>\chi^2_{\mathrm{th,3\sigma}})\simeq 82\%$. This confirms that the significance measured for the reported Crab profile is fully consistent with the expected non-central $\chi^2$ distribution in see \autoref{appendix:Chi_distribution}.

The consistency between the analytical/MonteCarlo predictions and the observed data validates the instrument performance, indicating that the results are governed by Poisson statistics.

\section{Discussion and Conclusion}

We performed millisecond X‑ray timing with an on‑axis dataset of $T_{\exp}=730\,$s from a single 1200~s operation, obtained by the wide field of view \emph{HERMES} payload onboard the 6U $\sim$11 kg nanosatellite \emph{SpIRIT}. By folding photon time of arrivals at the JB ephemeris, extrapolated to the observation epoch, we resolved the canonical Crab pulsar double‑peaked pulse profile with a statistical significance of $5.0\sigma$ in the $3–11.5$ keV band. The \emph{SpIRIT/HERMES} pulse shape was validated through a template fit to the closest-in-time NuSTAR observation of the Crab pulsar.
Moreover, an epoch‑folding search on the SpIRIT data, recovered a best period of $33.839$ ms, only $0.5\,\mu$s shorter than the expected value of the catalog and entirely consistent with the Fourier resolution limit of $\Delta P\sim 1.5~\mu$s achievable with the available \emph{SpIRIT/HERMES} exposure time.

These results demonstrate that the instrument atomic clock, the \textsc{SGP4} orbit propagation, and the \texttt{barycorr} correction, preserve the phase coherence to within a few microseconds over the 20‑minute observation window. 

In addition, to confirm the validity of the \emph{SpIRIT} pulse profile significance, we carried out an analytical treatment and Monte Carlo simulations using the \emph{RXTE} Crab pulse profile as a theoretical template. This allowed us to predict the expected significance by taking into account the exposure time and the characteristics of the \emph{HERMES} payload (e.g., instrument effective area, pointing, response matrix) for the Crab observation. Since the computed significance is in full agreement with these predictions, we can confirm that the instrument is operating correctly and that the measurement process is governed by the expected Poisson statistics.


Our conclusion is that high-precision timing, once restricted to large, resource-intensive observatories, can be achieved with cost-effective CubeSats. 
Reaching a $\gtrsim5\sigma$ detection of a short-period pulsar in just 12 minutes of exposure, with a $\lesssim50$cm$^{2}$ instrument effective area, highlights both the effectiveness of the highly compact \emph{HERMES} payload and the potential of low-cost, fast-development nanosatellite missions for high-energy astrophysics.

\begin{acknowledgments}
This work was supported by the Horizon 2020 programme (AHEAD2020, Grant Agreement No. 871158; HERMES-SP, Grant Agreement No. 821896), and from ASI–INAF agreements for the HERMES Technological Pathfinder (HTP; scientific activities) and HERMES Pathfinder (operations and scientific exploitation). The \emph{SpIRIT} mission is led by the University of Melbourne (PI Michele Trenti), which received funding through the International Space Investment - Expand Capability grant funding scheme managed by the ASA for mission development and launch, and through the Moon to Mars - Mission Demonstrator scheme of ASA for mission operations. WL acknowledges doctoral funding from the Space Science and Technology PhD programme (University of Trento, Cycle XXXVIII). MD acknowledges support from the Czech Science Foundation (GAČR), project No. 24-11487J.
\end{acknowledgments}

\bibliography{sample701}{}

@inproceedings{Campana_2020,
   title={The HERMES-TP/SP background and response simulations},
   url={http://dx.doi.org/10.1117/12.2560365},
   DOI={10.1117/12.2560365},
   booktitle={Space Telescopes and Instrumentation 2020: Ultraviolet to Gamma Ray},
   publisher={SPIE},
   author={Campana, Riccardo and Fuschino, Fabio and Evangelista, Yuri and Dilillo, Giuseppe and Fiore, Fabrizio},
   editor={den Herder, Jan-Willem A. and Nakazawa, Kazuhiro and Nikzad, Shouleh},
   year={2020},
   month=dec }

@article{gandola2021multi,
  title={Multi-chip front-end electronics LYRA for X and $\gamma$ Ray detector for HERMES mission},
  author={Gandola, M and Mele, F and Grassi, M and Malcovati, P and Bertuccio, G},
  journal={Journal of Instrumentation},
  volume={16},
  number={12},
  pages={T12013},
  year={2021},
  publisher={IOP Publishing},
  doi={10.1088/1748-0221/16/12/T12013}
}

@ARTICLE{1999ApJ...520..124G,
       author = {{Gruber}, D.~E. and {Matteson}, J.~L. and {Peterson}, L.~E. and {Jung}, G.~V.},
        title = "{The Spectrum of Diffuse Cosmic Hard X-Rays Measured with HEAO 1}",
      journal = {\apj},
         year = 1999,
        month = jul,
       volume = {520},
       number = {1},
        pages = {124-129},
          doi = {10.1086/307450},
archivePrefix = {arXiv},
       eprint = {astro-ph/9903492},
 primaryClass = {astro-ph},
       adsurl = {https://ui.adsabs.harvard.edu/abs/1999ApJ...520..124G}
}

@INPROCEEDINGS{Baroni2024,
       author = {{Baroni}, G. and {Campana}, R. and {Evangelista}, Y. and {Guzm{\'a}n}, A. and {Hedderman}, P. and {Dilillo}, G. and {Marchesini}, E.~J. and {Della Casa}, G. and {Trenti}, M. and {McRobbie}, J. and {Mearns}, R. and {Ortiz Del Castillo}, M. and {Thomas}, M. and {Therakam}, C. and {Chapman}, A. and {Pirrotta}, S. and {Puccetti}, S. and {Perri}, M. and {Santangelo}, A. and {Leone}, W. and {Trevisan}, S. and {Citossi}, M. and {Fiore}, F.},
        title = "{The commissioning and early operations of the high-energy HERMES payload onboard SpIRIT}",
    booktitle = {Space Telescopes and Instrumentation 2024: Ultraviolet to Gamma Ray},
         year = 2024,
       editor = {{den Herder}, Jan-Willem A. and {Nikzad}, Shouleh and {Nakazawa}, Kazuhiro},
       series = {Society of Photo-Optical Instrumentation Engineers (SPIE) Conference Series},
       volume = {13093},
        month = aug,
          eid = {130935O},
        pages = {130935O},
          doi = {10.1117/12.3018308},
       adsurl = {https://ui.adsabs.harvard.edu/abs/2024SPIE13093E..5OB}
}

@inproceedings{Evangelista_2024,
   title={The HERMES (High Energy Rapid Modular Ensemble of Satellites) Pathfinder mission},
   url={http://dx.doi.org/10.1117/12.3018259},
   DOI={10.1117/12.3018259},
   booktitle={Space Telescopes and Instrumentation 2024: Ultraviolet to Gamma Ray},
   publisher={SPIE},
   author={Evangelista, Yuri and Fiore, Fabrizio and Campana, Riccardo and Baroni, Giulia and Ceraudo, Francesco and Della Casa, Giovanni and Demenev, Evgeny and Dilillo, Giuseppe and Fiorini, Mauro and Ghirlanda, Giancarlo and Grassi, Marco and Guzmán Cabrera, Alejandro and Hedderman, Paul and Marchesini, Ezequiel J. and Morgante, Gianluca and Mele, Filippo and Nava, Lara and Nogara, Paolo and Nuti, Alessio and Pliego-Caballero, Samuel and Rashevskaya, Irina and Russo, Francesco and Sottile, Giuseppe and Lavagna, Michèle and Colagrossi, Andrea and Silvestrini, Stefano and Quirino, Matteo and Bechini, Michele and Brandonisio, Andrea and De Cecio, Francesco and Dottori, Alice and Troisi, Ivan and Bertacin, Roberto and Bellutti, Pierluigi and Bertuccio, Giuseppe and Burderi, Luciano and Chen, Tianxiang and Citossi, Marco and Di Salvo, Tiziana and Feroci, Marco and Ficorella, Francesco and Gao, Na and Grappasonni, Chiara and Labanti, Claudio and La Rosa, Giovanni and Leone, Wladimiro and Malcovati, Piero and Negri, Barbara and Pepponi, Giancarlo and Perri, Matteo and Piazzolla, Raffaele and Picciotto, Antonino and Pirrotta, Simone and Puccetti, Simonetta and Rashevsky, Alexander and Riggio, Alessandro and Rinaldi, Marianna and Sanna, Andrea and Santangelo, Andrea and Tenzer, Christoph and Tiberia, Alessandra and Trenti, Michele and Trevisan, Sara and Vacchi, Andrea and Xiong, Shaolin and Zampa, Gianluigi and Zampa, Nicola and Zhang, Shuangnan and Zorzi, Nicola and Ripa, Jakub and Werner, Norbert},
   editor={den Herder, Jan-Willem A. and Nakazawa, Kazuhiro and Nikzad, Shouleh},
   year={2024},
   month=aug, pages={74} }

@ARTICLE{ortizdelcastillo2025b_themis,
       author = {{Ortiz del Castillo}, Miguel and {Therakam}, Clint and {McRobbie}, Jack and {Woods}, Andrew and {Mearns}, Robert and {Barraclough}, Simon and {Catsamas}, Stephen and {Ohkawa}, Mika and {Morgan}, Jonathan and {Chapman}, Airlie and {Trenti}, Michele},
        title = "{Unlocking the Potential of Small Satellites: TheMIS's Active Cooling Technology on the SpIRIT Mission}",
      journal = {arXiv e-prints},
         year = 2024,
        month = jul,
          eid = {arXiv:2407.14031},
        pages = {arXiv:2407.14031},
          doi = {10.48550/arXiv.2407.14031},
archivePrefix = {arXiv},
       eprint = {2407.14031},
 primaryClass = {astro-ph.IM},
       adsurl = {https://ui.adsabs.harvard.edu/abs/2024arXiv240714031O}
}

@ARTICLE{Mearns2024_Mercury,
       author = {{Mearns}, Robert and {Chapman}, Airlie and {Trenti}, Michele},
        title = "{Characterisation of SATCOM Networks for Rapid Message Delivery: Early In-Orbit Results}",
      journal = {arXiv e-prints},
         year = 2024,
        month = jul,
          eid = {arXiv:2407.19623},
        pages = {arXiv:2407.19623},
          doi = {10.48550/arXiv.2407.19623},
archivePrefix = {arXiv},
       eprint = {2407.19623},
 primaryClass = {astro-ph.IM},
       adsurl = {https://ui.adsabs.harvard.edu/abs/2024arXiv240719623M}
}

@article{OrtizDelCastillo2025_PMS,
title = {The SpIRIT Payload Management System: A paradigm for smallsats multi payload integration},
journal = {Acta Astronautica},
volume = {236},
pages = {359-370},
year = {2025},
issn = {0094-5765},
doi = {https://doi.org/10.1016/j.actaastro.2025.06.050},
url = {https://www.sciencedirect.com/science/article/pii/S0094576525004023},
author = {Miguel Ortiz {del Castillo} and Jack McRobbie and Andrew Woods and Robert Mearns and Jonathan Morgan and Michele Trenti and Simon Barraclough and Yijie Tao and Clint Therakam and Eric Schoof and Airlie Chapman and Fabrizio Fiore and Yuri Evangelista and Alejandro Guzman and Paul Hedderman}
}

@ARTICLE{ortizdelcastillo2024a_loris,
       author = {{Ortiz del Castillo}, Miguel and {Morgan}, Jonathan and {McRobbie}, Jack and {Therakam}, Clint and {Joukhadar}, Zaher and {Mearns}, Robert and {Barraclough}, Simon and {Sinnott}, Richard and {Woods}, Andrew and {Bayliss}, Chris and {Ehinger}, Kris and {Rubinstein}, Ben and {Bailey}, James and {Chapman}, Airlie and {Trenti}, Michele},
        title = "{Mitigating Challenges of the Space Environment for Onboard Artificial Intelligence: Design Overview of the Imaging Payload on SpIRIT}",
      journal = {arXiv e-prints},
         year = 2024,
        month = apr,
          eid = {arXiv:2404.08399},
        pages = {arXiv:2404.08399},
          doi = {10.48550/arXiv.2404.08399},
archivePrefix = {arXiv},
       eprint = {2404.08399},
 primaryClass = {cs.CV},
       adsurl = {https://ui.adsabs.harvard.edu/abs/2024arXiv240408399O}
}

@misc{Fiore2025,
      title={HERMES Pathfinder \& SpIRIT: a progress report}, 
      author={F. Fiore and M. Trenti and Y. Evangelista and R. Campana and G. Baroni and F. Ceraudo and M. Citossi and G. Della Casa and G. Dilillo and M. Feroci and M. Fiorini and G. Ghirlanda and C. Labanti and G. La Rosa and E. J. Marchesini and G. Morgante and L. Nava and P. Nogara and A. Nuti and M. Perri and F. Russo and G. Sottile and M. Lavagna. A. Colagrossi and S. Silvestrini and M. Quirino and M. Bechini and L. Bianchi and A. Brandonisio and F. De Cecio and A. Dottori and I. Troisi and G. Bertuccio and F. Mele and B. Negri and R. Bertacin and C. Grappasonni and R. Piazzolla and S. Pirrotta and S. Puccetti and M. Rinaldi and A. Tiberia and L. Burderi and A. Sanna and A. Riggio and C. Cabras and A. Tsvetkova and A. Santangelo and A. Guzman and P. Hedderman and S. Pliego Cagallero and C. Tenzer and A. Vacchi and N. Zampa and R. Crupi and P. Bellutti and E. Demenev and F. Ficorella and D. Novel and G. Pepponi and A. Picciotto and N. Zorzi and M. Grassi and P. Malcovati and T. Di Salvo and W. Leone and S. Trevisan and I Rashevskaya and A. Rachevski and G. Zampa and T. Chen and N. Gao and S. Xiong and S. Yi and S. Zhang and M. Ortiz del Castillo and R. Mearns and J. McRobbie and A. Chapman and M. Thomas and A. Woods and J. Morgan and S. Barraclough and N. Werner and J. Ripa and F. Munz and A. Pal and D. Gacnik and A. Hudrap and D. Selkan and G. Molera Calves},
      year={2025},
      eprint={2502.17952},
      archivePrefix={arXiv},
      primaryClass={astro-ph.IM},
      url={https://arxiv.org/abs/2502.17952}, 
}

@misc{fiore2021distributed,
      title={Distributed Architectures and Constellations for Gamma-Ray Burst Science}, 
      author={F. Fiore and N. Werner and E. Behar},
      year={2021},
      eprint={2112.08982},
      archivePrefix={arXiv},
      primaryClass={astro-ph.HE}
}

@inproceedings{Fiore_2020,
   title={The HERMES-technologic and scientific pathfinder},
   url={http://dx.doi.org/10.1117/12.2560680},
   DOI={10.1117/12.2560680},
   booktitle={Space Telescopes and Instrumentation 2020: Ultraviolet to Gamma Ray},
   publisher={SPIE},
   author={Fiore, Fabrizio and Burderi, Luciano and Lavagna, Michelle and Bertacin, Roberto and Evangelista, Yuri and Campana, Riccardo and Fuschino, Fabio and Lunghi, Paolo and Monge, Angel and Negri, Barbara and Pirrotta, Simone and Puccetti, Simonetta and Sanna, Andrea and Amarilli, Fabrizio and Ambrosino, Filippo and Amelino-Camelia, Giovanni and Anitra, Alessio and Auricchio, Natalia and Barbera, Marco and Bechini, Michele and Bellutti, Pierluigi and Bertuccio, Giuseppe and Cao, Jiewei and Ceraudo, Francesco and Chen, Tianxiang and Cinelli, Marco and Citossi, Marco and Clerici, Aurora and Colagrossi, Andrea and Curzel, Serena and Della Casa, Giovanni and Demenev, Evgeny and Del Santo, Melania and Dilillo, Giuseppe and Di Salvo, Tiziana and Efremov, Pavel and Feroci, Marco and Feruglio, Chiara and Ferrandi, Fabrizio and Fiorini, Mauro and Fiorito, Michele and Frontera, Filippo and Gacnik, Dejan and Galgoczi, Gabor and Gao, Na and Gambino, Angelo Francesco and Gandola, Massimo and Ghirlanda, Giancarlo and Gomboc, Andreja and Grassi, Marco and Guzman, Alejandro and Karlica, Mile and Kostic, Uros and Labanti, Claudio and La Rosa, Giovanni and Lo Cicero, Ugo and Lopez-Fernandez, Borja and Malcovati, Piero and Maselli, Alessandro and Manca, Arianna and Mele, Filippo and Milankovich, Dorottya and Morgante, Gianluca and Nava, Lara and Nogara, Paolo and Ohno, Masanori and Ottolina, Daniele and Pasquale, Andrea and Pal, Andras and Perri, Matteo and Piazzolla, Raffaele and Piccinin, Margherita and Pliego-Caballero, Samuel and Prinetto, Jacopo and Pucacco, Giuseppe and Rashevsky, Alexander and Rashevskaya, Irina and Riggio, Alessandro and Ripa, Jakub and Russo, Francesco and Papitto, Alessandro and Piranomonte, Silvia and Santangelo, Andrea and Scala, Francesca and Sciarrone, Giulia and Selcan, David and Silvestrini, Stefano and Sottile, Giuseppe and Rotovnik, Tomaz and Tenzer, Christoph and Troisi, Ivan and Vacchi, Andrea and Virgilli, Enrico and Werner, Norbert and Wang, lingjun and Xu, Yupeng and Zampa, Gianluigi and Zampa, Nicola and Zanotti, Giovanni},
   editor={den Herder, Jan-Willem A. and Nakazawa, Kazuhiro and Nikzad, Shouleh},
   year={2020},
   month=dec }

@article{Ghirlanda_2024,
   title={HERMES: Gamma-ray burst and gravitational wave counterpart hunter},
   volume={689},
   ISSN={1432-0746},
   url={http://dx.doi.org/10.1051/0004-6361/202450006},
   DOI={10.1051/0004-6361/202450006},
   journal={Astronomy \&amp; Astrophysics},
   publisher={EDP Sciences},
   author={Ghirlanda, G. and Nava, L. and Salafia, O. and Fiore, F. and Campana, R. and Salvaterra, R. and Sanna, A. and Leone, W. and Evangelista, Y. and Dilillo, G. and Puccetti, S. and Santangelo, A. and Trenti, M. and Guzmán, A. and Hedderman, P. and Amelino-Camelia, G. and Barbera, M. and Baroni, G. and Bechini, M. and Bellutti, P. and Bertuccio, G. and Borghi, G. and Brandonisio, A. and Burderi, L. and Cabras, C. and Chen, T. and Citossi, M. and Colagrossi, A. and Crupi, R. and De Cecio, F. and Dedolli, I. and Del Santo, M. and Demenev, E. and Di Salvo, T. and Ficorella, F. and Gačnik, D. and Gandola, M. and Gao, N. and Gomboc, A. and Grassi, M. and Iaria, R. and La Rosa, G. and Lo Cicero, U. and Malcovati, P. and Manca, A. and Marchesini, E. J. and Maselli, A. and Mele, F. and Nogara, P. and Pepponi, G. and Perri, M. and Picciotto, A. and Pirrotta, S. and Prinetto, J. and Quirino, M. and Riggio, A. and Řípa, J. and Russo, F. and Selčan, D. and Silvestrini, S. and Sottile, G. and Thomas, M. L. and Tiberia, A. and Trevisan, S. and Troisi, I. and Tsvetkova, A. and Vacchi, A. and Werner, N. and Zanotti, G. and Zorzi, N.},
   year={2024},
   month=sep, pages={A175} }

@article{Jain_2011,
   title={Pulse profile stability of the Crab pulsar},
   volume={11},
   ISSN={1674-4527},
   url={http://dx.doi.org/10.1088/1674-4527/11/10/002},
   DOI={10.1088/1674-4527/11/10/002},
   number={10},
   journal={Research in Astronomy and Astrophysics},
   publisher={IOP Publishing},
   author={Jain, Chetana and Paul, Biswajit},
   year={2011},
   month=sep, pages={1134–1142} }

@inproceedings{Kirsch_2005,
   title={Crab: the standard x-ray candle with all (modern) x-ray satellites},
   ISSN={0277-786X},
   url={http://dx.doi.org/10.1117/12.616893},
   DOI={10.1117/12.616893},
   booktitle={UV, X-Ray, and Gamma-Ray Space Instrumentation for Astronomy XIV},
   publisher={SPIE},
   author={Kirsch, M. G. and Briel, U. G. and Burrows, D. and Campana, S. and Cusumano, G. and Ebisawa, K. and Freyberg, M. J. and Guainazzi, M. and Haberl, F. and Jahoda, K. and Kaastra, J. and Kretschmar, P. and Larsson, S. and Lubinski, P. and Mori, K. and Plucinsky, P. and Pollock, A. M. and Rothschild, R. and Sembay, S. and Wilms, J. and Yamamoto, M.},
   editor={Siegmund, Oswald H. W.},
   year={2005},
   month=aug }

@article{Weisskopf_2011,
   title={CHANDRAPHASE-RESOLVED X-RAY SPECTROSCOPY OF THE CRAB PULSAR},
   volume={743},
   ISSN={1538-4357},
   url={http://dx.doi.org/10.1088/0004-637X/743/2/139},
   DOI={10.1088/0004-637x/743/2/139},
   number={2},
   journal={The Astrophysical Journal},
   publisher={American Astronomical Society},
   author={Weisskopf, Martin C. and Tennant, Allyn F. and Yakovlev, Dmitry G. and Harding, Alice and Zavlin, Vyacheslav E. and O’Dell, Stephen L. and Elsner, Ronald F. and Becker, Werner},
   year={2011},
   month=nov, pages={139} }

@ARTICLE{Brandt2003,
       author = {{Brandt}, S. and {Budtz-J{\o}rgensen}, C. and {Lund}, N. and {Westergaard}, N.~J. and {Rasmussen}, I.~L. and {Andersen}, K.~H. and {Chenevez}, J. and {Hornstrup}, A. and {Jensen}, P.~A. and {Laursen}, S. and {Om{\o}}, K. and {Oxborrow}, C.~A. and {Pedersen}, S.~M. and {Polny}, J. and {Andersson}, H. and {Andersson}, T. and {Vilhu}, O. and {Huovelin}, J. and {Maisala}, S. and {Morawski}, M. and {Juchnikowski}, G. and {Costa}, E. and {Feroci}, M. and {Rubini}, A. and {Rapisarda}, M. and {Morelli}, E. and {Frontera}, F. and {Pelliciari}, C. and {Loffredo}, G. and {Carassiti}, V. and {Reglero}, V. and {Mart{\'\i}nez N{\'u}{\~n}ez}, S. and {Larsson}, S. and {Svensson}, R. and {Zdziarski}, A.~A. and {Castro-Tirado}, A. and {Goria}, M. and {Giulianelli}, G. and {Rezazad}, M. and {Cordero}, F. and {Schmidt}, M. and {Carli}, R. and {Jensen}, P.~L. and {Sarri}, G. and {Gomez}, C. and {Orr}, A. and {Much}, R. and {Schnopper}, H.~W. and {Kretschmar}, P.},
        title = "{JEM-X inflight performance}",
      journal = {\aap},
         year = 2003,
        month = nov,
       volume = {411},
        pages = {L243-L251},
          doi = {10.1051/0004-6361:20031363},
       adsurl = {https://ui.adsabs.harvard.edu/abs/2003A&A...411L.243B}
}

@misc{trenti2024,
      title={SpIRIT Mission: In-Orbit Results and Technology Demonstrations}, 
      author={Michele Trenti and Miguel Ortiz del Castillo and Robert Mearns and Jack McRobbie and Clint Therakam and Airlie Chapman and Andrew Woods and Jonathan Morgan and Simon Barraclough and Ivan Rodriguez Mallo and Giulia Baroni and Fabrizio Fiore and Yuri Evangelista and Riccardo Campana and Alejandro Guzman and Paul Hedderman},
      year={2024},
      eprint={2407.14034},
      archivePrefix={arXiv},
      primaryClass={astro-ph.IM},
      url={https://arxiv.org/abs/2407.14034}, 
}

@article{Leone_2025,
   title="{Time domain astrophysics with transient sources: Delay estimate via Cross Correlation Function techniques}",
   volume={701},
   ISSN={1432-0746},
   url={http://dx.doi.org/10.1051/0004-6361/202453054},
   DOI={10.1051/0004-6361/202453054},
   journal={Astronomy \&amp; Astrophysics},
   publisher={EDP Sciences},
   author={{Leone}, W. and {Burderi}, L. and {Di Salvo}, T. and {Anitra}, A. and {Sanna}, A. and {Riggio}, A. and {Iaria}, R. and {Fiore}, F. and {Longo}, F. and {Ďuríšková}, M. and {Tsvetkova}, A. and {Maraventano}, C. and {Miceli}, C.},
   year={2025},
   month=sep, 
pages={A50} }

@article{Thomas_2023,
   title={Localisation of gamma-ray bursts from the combined SpIRIT+HERMES-TP/SP nano-satellite constellation},
   volume={40},
   ISSN={1448-6083},
   url={http://dx.doi.org/10.1017/pasa.2023.4},
   DOI={10.1017/pasa.2023.4},
   journal={Publications of the Astronomical Society of Australia},
   publisher={Cambridge University Press (CUP)},
   author={Thomas, M. and Trenti, M. and Sanna, A. and Campana, R. and Ghirlanda, G. and Řípa, J. and Burderi, L. and Fiore, F. and Evangelista, Y. and Amati, L. and Barraclough, S. and Auchettl, K. and del Castillo, M. O. and Chapman, A. and Citossi, M. and Colagrossi, A. and Dilillo, G. and Deiosso, N. and Demenev, E. and Longo, F. and Marino, A. and McRobbie, J. and Mearns, R. and Melandri, A. and Riggio, A. and Di Salvo, T. and Puccetti, S. and Topinka, M.},
   year={2023} }

@article{Rots_2004,
   title={Absolute Timing of the Crab Pulsar with the
                    Rossi X-Ray Timing Explorer},
   volume={605},
   ISSN={1538-4357},
   url={http://dx.doi.org/10.1086/420842},
   DOI={10.1086/420842},
   number={2},
   journal={The Astrophysical Journal},
   publisher={American Astronomical Society},
   author={Rots, Arnold H. and Jahoda, Keith and Lyne, Andrew G.},
   year={2004},
   month=mar, pages={L129–L132} }

@inproceedings{Campana_2022,
   title={Calibration of the first detector flight models for the HERMES constellation and the SpIRIT mission},
   url={http://dx.doi.org/10.1117/12.2629031},
   DOI={10.1117/12.2629031},
   booktitle={Space Telescopes and Instrumentation 2022: Ultraviolet to Gamma Ray},
   publisher={SPIE},
   author={Campana, Riccardo and Baroni, Giulia and Della Casa, Giovanni and Dilillo, Giuseppe and Marchesini, Ezequiel J. and Ceraudo, Francesco and Guzmán, Alejandro and Hedderman, Paul and Evangelista, Yuri},
   editor={den Herder, Jan-Willem A. and Nakazawa, Kazuhiro and Nikzad, Shouleh},
   year={2022},
   month=aug, pages={249} }

@inproceedings{Campana_2024,
   title={Design and development of the HERMES Pathfinder payloads},
   url={http://dx.doi.org/10.1117/12.3018007},
   DOI={10.1117/12.3018007},
   booktitle={Space Telescopes and Instrumentation 2024: Ultraviolet to Gamma Ray},
   publisher={SPIE},
   author={Campana, Riccardo and Evangelista, Yuri and Fiore, Fabrizio and Guzmán Cabrera, Alejandro and Baroni, Giulia and Della Casa, Giovanni and Dilillo, Giuseppe and Hedderman, Paul and Marchesini, Ezequiel J. and Bertuccio, Giuseppe and Ceraudo, Francesco and Demenev, Evgeny and Fiorini, Mauro and Grassi, Marco and Malcovati, Piero and Mele, Filippo and Nogara, Paolo and Nuti, Alessio and Perri, Matteo and Pirrotta, Simone and Pliego-Caballero, Samuel and Puccetti, Simonetta and Sottile, Giuseppe and Russo, Francesco and Trevisan, Sara},
   editor={den Herder, Jan-Willem A. and Nakazawa, Kazuhiro and Nikzad, Shouleh},
   year={2024},
   month=aug, pages={245} }

@article{Dilillo_2024,
   title={The HERMES calibration pipeline: mescal},
   volume={46},
   ISSN={2213-1337},
   url={http://dx.doi.org/10.1016/j.ascom.2024.100797},
   DOI={10.1016/j.ascom.2024.100797},
   journal={Astronomy and Computing},
   publisher={Elsevier BV},
   author={Dilillo, G. and Marchesini, E.J. and Della Casa, G. and Baroni, G. and Campana, R. and Borciani, E. and Srivastava, S. and Trevisan, S. and Ceraudo, F. and Citossi, M. and Evangelista, Y. and Guzmán, A. and Hedderman, P. and Labanti, C. and Virgilli, E. and Fiore, F.},
   year={2024},
   month=jan, pages={100797} }

@misc{grbalphasmallestastrophysicalspace,
      title={GRBAlpha: the smallest astrophysical space observatory -- Part 1: Detector design, system description and satellite operations}, 
      author={András Pál and Masanori Ohno and László Mészáros and Norbert Werner and Jakub Řípa and Balázs Csák and Marianna Dafčíková and Marcel Frajt and Yasushi Fukazawa and Peter Hanák and Ján Hudec and Nikola Husáriková and Jakub Kapuš and Miroslav Kasal and Martin Kolář and Martin Koleda and Robert Laszlo and Pavol Lipovský and Tsunefumi Mizuno and Filip Münz and Kazuhiro Nakazawa and Maksim Rezenov and Miroslav Šmelko and Hiromitsu Takahashi and Martin Topinka and Tomáš Urbanec and Jean-Paul Breuer and Tamás Bozóki and Gergely Dálya and Teruaki Enoto and Zsolt Frei and Gergely Friss and Gábor Galgóczi and Filip Hroch and Yuto Ichinohe and Kornél Kapás and László L. Kiss and Hiroto Matake and Hirokazu Odaka and Helen Poon and Aleš Povalač and János Takátsy and Kento Torigoe and Nagomi Uchida and Yuusuke Uchida},
      year={2023},
      eprint={2302.10048},
      archivePrefix={arXiv},
      primaryClass={astro-ph.IM},
      url={https://arxiv.org/abs/2302.10048}, 
}

@article{Vivekanand_2021,
   title={Phase-resolved spectrum of the Crab pulsar from NICER},
   volume={649},
   ISSN={1432-0746},
   url={http://dx.doi.org/10.1051/0004-6361/202140358},
   DOI={10.1051/0004-6361/202140358},
   journal={Astronomy \&; Astrophysics},
   publisher={EDP Sciences},
   author={Vivekanand, M.},
   year={2021},
   month=may, pages={A140} }

@ARTICLE{Madsen_2015,
       author = {{Madsen}, Kristin K. and {Harrison}, Fiona A. and {Markwardt}, Craig B. and {An}, Hongjun and {Grefenstette}, Brian W. and {Bachetti}, Matteo and {Miyasaka}, Hiromasa and {Kitaguchi}, Takao and {Bhalerao}, Varun and {Boggs}, Steve and {Christensen}, Finn E. and {Craig}, William W. and {Forster}, Karl and {Fuerst}, Felix and {Hailey}, Charles J. and {Perri}, Matteo and {Puccetti}, Simonetta and {Rana}, Vikram and {Stern}, Daniel and {Walton}, Dominic J. and {J{\o}rgen Westergaard}, Niels and {Zhang}, William W.},
        title = "{Calibration of the NuSTAR High-energy Focusing X-ray Telescope.}",
      journal = {\apjs},
         year = 2015,
        month = sep,
       volume = {220},
       number = {1},
          eid = {8},
        pages = {8},
          doi = {10.1088/0067-0049/220/1/8},
archivePrefix = {arXiv},
       eprint = {1504.01672},
 primaryClass = {astro-ph.IM},
       adsurl = {https://ui.adsabs.harvard.edu/abs/2015ApJS..220....8M}
}

@INCOLLECTION{Klis_2006,
       author = {{van der Klis}, M.},
        title = "{Rapid X-ray Variability}",
    booktitle = {Compact stellar X-ray sources},
         year = 2006,
       editor = {{Lewin}, Walter H.~G. and {van der Klis}, Michiel},
       volume = {39},
        pages = {39-112},
       adsurl = {https://ui.adsabs.harvard.edu/abs/2006csxs.book...39V}
}

@ARTICLE{2013Hurley,
       author = {{Hurley}, K. and {Pal'shin}, V.~D. and {Aptekar}, R.~L. and {Golenetskii}, S.~V. and {Frederiks}, D.~D. and {Mazets}, E.~P. and {Svinkin}, D.~S. and {Briggs}, M.~S. and {Connaughton}, V. and {Meegan}, C. and {Goldsten}, J. and {Boynton}, W. and {Fellows}, C. and {Harshman}, K. and {Mitrofanov}, I.~G. and {Golovin}, D.~V. and {Kozyrev}, A.~S. and {Litvak}, M.~L. and {Sanin}, A.~B. and {Rau}, A. and {von Kienlin}, A. and {Zhang}, X. and {Yamaoka}, K. and {Fukazawa}, Y. and {Hanabata}, Y. and {Ohno}, M. and {Takahashi}, T. and {Tashiro}, M. and {Terada}, Y. and {Murakami}, T. and {Makishima}, K. and {Barthelmy}, S. and {Cline}, T. and {Gehrels}, N. and {Cummings}, J. and {Krimm}, H.~A. and {Smith}, D.~M. and {Del Monte}, E. and {Feroci}, M. and {Marisaldi}, M.},
        title = "{The Interplanetary Network Supplement to the Fermi GBM Catalog of Cosmic Gamma-Ray Bursts}",
      journal = {\apjs},
         year = 2013,
        month = aug,
       volume = {207},
       number = {2},
          eid = {39},
        pages = {39},
          doi = {10.1088/0067-0049/207/2/39},
archivePrefix = {arXiv},
       eprint = {1301.3522},
 primaryClass = {astro-ph.HE},
       adsurl = {https://ui.adsabs.harvard.edu/abs/2013ApJS..207...39H}
}

@ARTICLE{1999Swank,
       author = {{Swank}, J.~H.},
        title = "{The Rossi X-Ray Timing Explorer}",
      journal = {Nuclear Physics B Proceedings Supplements},
         year = 1999,
        month = jan,
       volume = {69},
       number = {1-3},
        pages = {12-19},
          doi = {10.1016/S0920-5632(98)00175-3},
archivePrefix = {arXiv},
       eprint = {astro-ph/9802188},
 primaryClass = {astro-ph},
       adsurl = {https://ui.adsabs.harvard.edu/abs/1999NuPhS..69...12S}
}

@inbook{Vallado2006,
author = {David Vallado and Paul Crawford and Ricahrd Hujsak and T.S. Kelso},
title = {Revisiting Spacetrack Report \#3},
booktitle = {AIAA/AAS Astrodynamics Specialist Conference and Exhibit},
chapter = {},
pages = {1-94},
year = {2006},
publisher = {AIAA},

doi = {10.2514/6.2006-6753},
URL = {https://arc.aiaa.org/doi/abs/10.2514/6.2006-6753},
eprint = {https://arc.aiaa.org/doi/pdf/10.2514/6.2006-6753}
}

@article{Tamagawa_2025,
   title={NinjaSat: Astronomical X-ray CubeSat observatory},
   volume={77},
   ISSN={2053-051X},
   url={http://dx.doi.org/10.1093/pasj/psaf014},
   DOI={10.1093/pasj/psaf014},
   number={3},
   journal={Publications of the Astronomical Society of Japan},
   publisher={Oxford University Press (OUP)},
   author={Tamagawa, Toru and Enoto, Teruaki and Kitaguchi, Takao and Iwakiri, Wataru and Kato, Yo and Numazawa, Masaki and Mihara, Tatehiro and Takeda, Tomoshi and Ota, Naoyuki and Watanabe, Sota and Aoyama, Amira and Iwata, Satoko and Takahashi, Takuya and Yamasaki, Kaede and Hu, Chin-Ping and Takahashi, Hiromitsu and Yoshida, Yuto and Sato, Hiroki and Hayashi, Shoki and Zhou, Yuanhui and Uchiyama, Keisuke and Jujo, Arata and Odaka, Hirokazu and Tamba, Tsubasa and Taniguchi, Kentaro},
   year={2025},
   month=may, pages={466–479} }

@inproceedings{Evangelista2024,
   title={The scientific payload on-board the HERMES-TP and HERMES-SP CubeSat missions},
   url={http://dx.doi.org/10.1117/12.2561018},
   DOI={10.1117/12.2561018},
   booktitle={Space Telescopes and Instrumentation 2020: Ultraviolet to Gamma Ray},
   publisher={SPIE},
   author={Evangelista, Yuri and Fiore, Fabrizio and Fuschino, Fabio and Campana, Riccardo and Ceraudo, Francesco and Demenev, Evgeny and Guzman, Alejandro and Labanti, Claudio and La Rosa, Giovanni and Fiorini, Mauro and Gandola, Massimo and Grassi, Marco and Mele, Filippo and Morgante, Gianluca and Nogara, Paolo and Piazzolla, Raffaele and Pliego Caballero, Samuel and Rashevskaya, Irina and Russo, Francesco and Sciarrone, Giulia and Sottile, Giuseppe and Milankovich, Dorottya and Pál, András and Ambrosino, Filippo and Auricchio, Natalia and Barbera, Marco and Bellutti, Pierluigi and Bertuccio, Giuseppe and Borghi, Giacomo and Cao, Jiewei and Chen, Tianxiang and Dilillo, Giuseppe and Feroci, Marco and Ficorella, Francesco and Lo Cicero, Ugo and Malcovati, Piero and Morbidini, Alfredo and Pauletta, Giovanni and Picciotto, Antonino and Rachevski, Alexandre and Santangelo, Andrea and Tenzer, Chistoph and Vacchi, Andrea and Wang, Lingjun and Xu, Yupeng and Zampa, Gianluigi and Zampa, Nicola and Zorzi, Nicola and Burderi, Luciano and Lavagna, Michèle and Bertacin, Roberto and Lunghi, Paolo and Monge, Angel and Negri, Barbara and Pirrotta, Simone and Puccetti, Simonetta and Sanna, Andrea and Amarilli, Fabrizio and Amelino-Camelia, Giovanni and Bechini, Michele and Citossi, Marco and Colagrossi, Andrea and Curzel, Serena and Della Casa, Giovanni and Cinelli, Marco and Del Santo, Melania and Di Salvo, Tiziana and Feruglio, Chiara and Ferrandi, Fabrizio and Fiorito, Michele and Gacnik, Dejan and Galgóczi, Gabor and Gambino, Angelo Francesco and Ghirlanda, Giancarlo and Gomboc, Andreja and Karlica, Mile and Efremov, Pavel and Kostic, Uros and Clerici, Aurora and Lopez Fernandez, Borja and Maselli, Alessandro and Nava, Lara and Ohno, Masanori and Ottolina, Daniele and Pasquale, Andrea and Perri, Matteo and Piccinin, Margherita and Prinetto, Jacopo and Riggio, Alessandro and Ripa, Jakub and Papitto, Alessandro and Piranomonte, Silvia and Scala, Francesca and Selcan, David and Silvestrini, Stefano and Rotovnik, Tomaz and Virgilli, Enrico and Troisi, Ivan and Werner, Norbert and Zanotti, Giovanni and Anitra, Alessio and Manca, Arianna and Clerici, Aurora},
   editor={den Herder, Jan-Willem A. and Nakazawa, Kazuhiro and Nikzad, Shouleh},
   year={2020},
   month=dec }

@ARTICLE{GaiaDR3,
       author = {{Gaia Collaboration} and {Vallenari}, A. and {Brown}, A.~G.~A. and {Prusti}, T. and {de Bruijne}, J.~H.~J. and {Arenou}, F. and {Babusiaux}, C. and {Biermann}, M. and {Creevey}, O.~L. and {Ducourant}, C. and {Evans}, D.~W. and {Eyer}, L. and {Guerra}, R. and {Hutton}, A. and {Jordi}, C. and {Klioner}, S.~A. and {Lammers}, U.~L. and {Lindegren}, L. and {Luri}, X. and {Mignard}, F. and {Panem}, C. and {Pourbaix}, D. and {Randich}, S. and {Sartoretti}, P. and {Soubiran}, C. and {Tanga}, P. and {Walton}, N.~A. and {Bailer-Jones}, C.~A.~L. and {Bastian}, U. and {Drimmel}, R. and {Jansen}, F. and {Katz}, D. and {Lattanzi}, M.~G. and {van Leeuwen}, F. and {Bakker}, J. and {Cacciari}, C. and {Casta{\~n}eda}, J. and {De Angeli}, F. and {Fabricius}, C. and {Fouesneau}, M. and {Fr{\'e}mat}, Y. and {Galluccio}, L. and {Guerrier}, A. and {Heiter}, U. and {Masana}, E. and {Messineo}, R. and {Mowlavi}, N. and {Nicolas}, C. and {Nienartowicz}, K. and {Pailler}, F. and {Panuzzo}, P. and {Riclet}, F. and {Roux}, W. and {Seabroke}, G.~M. and {Sordo}, R. and {Th{\'e}venin}, F. and {Gracia-Abril}, G. and {Portell}, J. and {Teyssier}, D. and {Altmann}, M. and {Andrae}, R. and {Audard}, M. and {Bellas-Velidis}, I. and {Benson}, K. and {Berthier}, J. and {Blomme}, R. and {Burgess}, P.~W. and {Busonero}, D. and {Busso}, G. and {C{\'a}novas}, H. and {Carry}, B. and {Cellino}, A. and {Cheek}, N. and {Clementini}, G. and {Damerdji}, Y. and {Davidson}, M. and {de Teodoro}, P. and {Nu{\~n}ez Campos}, M. and {Delchambre}, L. and {Dell'Oro}, A. and {Esquej}, P. and {Fern{\'a}ndez-Hern{\'a}ndez}, J. and {Fraile}, E. and {Garabato}, D. and {Garc{\'\i}a-Lario}, P. and {Gosset}, E. and {Haigron}, R. and {Halbwachs}, J.-L. and {Hambly}, N.~C. and {Harrison}, D.~L. and {Hern{\'a}ndez}, J. and {Hestroffer}, D. and {Hodgkin}, S.~T. and {Holl}, B. and {Jan{\ss}en}, K. and {Jevardat de Fombelle}, G. and {Jordan}, S. and {Krone-Martins}, A. and {Lanzafame}, A.~C. and {L{\"o}ffler}, W. and {Marchal}, O. and {Marrese}, P.~M. and {Moitinho}, A. and {Muinonen}, K. and {Osborne}, P. and {Pancino}, E. and {Pauwels}, T. and {Recio-Blanco}, A. and {Reyl{\'e}}, C. and {Riello}, M. and {Rimoldini}, L. and {Roegiers}, T. and {Rybizki}, J. and {Sarro}, L.~M. and {Siopis}, C. and {Smith}, M. and {Sozzetti}, A. and {Utrilla}, E. and {van Leeuwen}, M. and {Abbas}, U. and {{\'A}brah{\'a}m}, P. and {Abreu Aramburu}, A. and {Aerts}, C. and {Aguado}, J.~J. and {Ajaj}, M. and {Aldea-Montero}, F. and {Altavilla}, G. and {{\'A}lvarez}, M.~A. and {Alves}, J. and {Anders}, F. and {Anderson}, R.~I. and {Anglada Varela}, E. and {Antoja}, T. and {Baines}, D. and {Baker}, S.~G. and {Balaguer-N{\'u}{\~n}ez}, L. and {Balbinot}, E. and {Balog}, Z. and {Barache}, C. and {Barbato}, D. and {Barros}, M. and {Barstow}, M.~A. and {Bartolom{\'e}}, S. and {Bassilana}, J.-L. and {Bauchet}, N. and {Becciani}, U. and {Bellazzini}, M. and {Berihuete}, A. and {Bernet}, M. and {Bertone}, S. and {Bianchi}, L. and {Binnenfeld}, A. and {Blanco-Cuaresma}, S. and {Blazere}, A. and {Boch}, T. and {Bombrun}, A. and {Bossini}, D. and {Bouquillon}, S. and {Bragaglia}, A. and {Bramante}, L. and {Breedt}, E. and {Bressan}, A. and {Brouillet}, N. and {Brugaletta}, E. and {Bucciarelli}, B. and {Burlacu}, A. and {Butkevich}, A.~G. and {Buzzi}, R. and {Caffau}, E. and {Cancelliere}, R. and {Cantat-Gaudin}, T. and {Carballo}, R. and {Carlucci}, T. and {Carnerero}, M.~I. and {Carrasco}, J.~M. and {Casamiquela}, L. and {Castellani}, M. and {Castro-Ginard}, A. and {Chaoul}, L. and {Charlot}, P. and {Chemin}, L. and {Chiaramida}, V. and {Chiavassa}, A. and {Chornay}, N. and {Comoretto}, G. and {Contursi}, G. and {Cooper}, W.~J. and {Cornez}, T. and {Cowell}, S. and {Crifo}, F. and {Cropper}, M. and {Crosta}, M. and {Crowley}, C. and {Dafonte}, C. and {Dapergolas}, A. and {David}, M. and {David}, P. and {de Laverny}, P. and {De Luise}, F. and {De March}, R.},
        title = "{Gaia Data Release 3. Summary of the content and survey properties}",
      journal = {\aap},
         year = 2023,
        month = jun,
       volume = {674},
          eid = {A1},
        pages = {A1},
          doi = {10.1051/0004-6361/202243940},
archivePrefix = {arXiv},
       eprint = {2208.00211},
 primaryClass = {astro-ph.GA},
       adsurl = {https://ui.adsabs.harvard.edu/abs/2023A&A...674A...1G}
}

@INPROCEEDINGS{Puccetti2024,
       author = {{Puccetti}, S. and {Perri}, M. and {Campana}, R. and {Marchesini}, E. and {Baroni}, G. and {Dilillo}, G. and {Evangelista}, Y. and {Sanna}, A. and {Burderi}, L. and {Fiore}, F.},
        title = "{HERMES SOC activities at the ASI space science data center (SSDC)}",
    booktitle = {Observatory Operations: Strategies, Processes, and Systems X},
         year = 2024,
       editor = {{Benn}, Chris R. and {Chrysostomou}, Antonio and {Storrie-Lombardi}, Lisa J.},
       series = {Society of Photo-Optical Instrumentation Engineers (SPIE) Conference Series},
       volume = {13098},
        month = jul,
          eid = {130980O},
        pages = {130980O},
          doi = {10.1117/12.3018828},
       adsurl = {https://ui.adsabs.harvard.edu/abs/2024SPIE13098E..0OP}
}

@ARTICLE{Scargle1982,
       author = {{Scargle}, J.~D.},
        title = "{Studies in astronomical time series analysis. II. Statistical aspects of spectral analysis of unevenly spaced data.}",
      journal = {\apj},
         year = 1982,
        month = dec,
       volume = {263},
        pages = {835-853},
          doi = {10.1086/160554},
       adsurl = {https://ui.adsabs.harvard.edu/abs/1982ApJ...263..835S}
}

@ARTICLE{Lomb1976,
       author = {{Lomb}, N.~R.},
        title = "{Least-Squares Frequency Analysis of Unequally Spaced Data}",
      journal = {\apss},
         year = 1976,
        month = feb,
       volume = {39},
       number = {2},
        pages = {447-462},
          doi = {10.1007/BF00648343},
       adsurl = {https://ui.adsabs.harvard.edu/abs/1976Ap&SS..39..447L}
}

@inproceedings{Evangelista_2020,
   title={The scientific payload on-board the HERMES-TP and HERMES-SP CubeSat missions},
   url={http://dx.doi.org/10.1117/12.2561018},
   DOI={10.1117/12.2561018},
   booktitle={Space Telescopes and Instrumentation 2020: Ultraviolet to Gamma Ray},
   publisher={SPIE},
   author={Evangelista, Yuri and Fiore, Fabrizio and Fuschino, Fabio and Campana, Riccardo and Ceraudo, Francesco and Demenev, Evgeny and Guzman, Alejandro and Labanti, Claudio and La Rosa, Giovanni and Fiorini, Mauro and Gandola, Massimo and Grassi, Marco and Mele, Filippo and Morgante, Gianluca and Nogara, Paolo and Piazzolla, Raffaele and Pliego Caballero, Samuel and Rashevskaya, Irina and Russo, Francesco and Sciarrone, Giulia and Sottile, Giuseppe and Milankovich, Dorottya and Pál, András and Ambrosino, Filippo and Auricchio, Natalia and Barbera, Marco and Bellutti, Pierluigi and Bertuccio, Giuseppe and Borghi, Giacomo and Cao, Jiewei and Chen, Tianxiang and Dilillo, Giuseppe and Feroci, Marco and Ficorella, Francesco and Lo Cicero, Ugo and Malcovati, Piero and Morbidini, Alfredo and Pauletta, Giovanni and Picciotto, Antonino and Rachevski, Alexandre and Santangelo, Andrea and Tenzer, Chistoph and Vacchi, Andrea and Wang, Lingjun and Xu, Yupeng and Zampa, Gianluigi and Zampa, Nicola and Zorzi, Nicola and Burderi, Luciano and Lavagna, Michèle and Bertacin, Roberto and Lunghi, Paolo and Monge, Angel and Negri, Barbara and Pirrotta, Simone and Puccetti, Simonetta and Sanna, Andrea and Amarilli, Fabrizio and Amelino-Camelia, Giovanni and Bechini, Michele and Citossi, Marco and Colagrossi, Andrea and Curzel, Serena and Della Casa, Giovanni and Cinelli, Marco and Del Santo, Melania and Di Salvo, Tiziana and Feruglio, Chiara and Ferrandi, Fabrizio and Fiorito, Michele and Gacnik, Dejan and Galgóczi, Gabor and Gambino, Angelo Francesco and Ghirlanda, Giancarlo and Gomboc, Andreja and Karlica, Mile and Efremov, Pavel and Kostic, Uros and Clerici, Aurora and Lopez Fernandez, Borja and Maselli, Alessandro and Nava, Lara and Ohno, Masanori and Ottolina, Daniele and Pasquale, Andrea and Perri, Matteo and Piccinin, Margherita and Prinetto, Jacopo and Riggio, Alessandro and Ripa, Jakub and Papitto, Alessandro and Piranomonte, Silvia and Scala, Francesca and Selcan, David and Silvestrini, Stefano and Rotovnik, Tomaz and Virgilli, Enrico and Troisi, Ivan and Werner, Norbert and Zanotti, Giovanni and Anitra, Alessio and Manca, Arianna and Clerici, Aurora},
   editor={den Herder, Jan-Willem A. and Nakazawa, Kazuhiro and Nikzad, Shouleh},
   year={2020},
   month=dec }
\bibliographystyle{aasjournalv7}
    
\appendix
\label{appendix}

\section{Crab Pulsar Exposure with HERMES: Analytical Estimate with MC Validation}

In general, determining the exposure time required to detect a neutron star pulsed profile is not straightforward.
The calculations presented in this appendix show, both analytically and through simulations, how to assess an $n\sigma$-level detection as a function of a given exposure time, $T_{\mathrm{exp}}$.

We adopt as a template the folded \emph{Rossi X-ray Timing Explorer} \citep[RXTE,][]{1999Swank} pulse profile of the Crab pulsar reported by \cite{Jain_2011}. Assuming an on–axis pointing, we compute the exposure time required for the two peaks to reach the desired statistical significance.

By considering the detector response files, we simulated the count spectrum expected for \emph{HERMES} with the \texttt{fakeit} task in \textsc{xspec}, modeling it as the superposition of three components:  
(i) the Crab Nebula, whose spectral parameters were taken from \citet{Kirsch_2005};  
(ii) the pulsed emission of the Crab pulsar, described by the model in \citet{Weisskopf_2011}; and  
(iii) the diffuse X‑ray background.

The CXB is derived from the diffused CXB model proposed in \cite{1999ApJ...520..124G}. We assume a low energy \emph{HERMES}/\emph{SpIRIT} FoV of 1.57 steradians as in \cite{Ghirlanda_2024}, for the 3--11.5 keV energy band.

In the current configuration of the \emph{HERMES} instrument onboard \emph{SpIRIT}, the nominal energy lower cut threshold is set at a level equivalent to about 5~keV. However, the precise energy threshold for each active channel may vary due to channel-to-channel process variations and the discrimination circuit's working principle \cite{Evangelista2024}. This leads to a significant threshold spread, and a particular sensitivity also in the 3--5~keV band. 
Accordingly, throughout we adopted a lower-energy cut of 3~keV.

An appropriate set of \emph{HERMES} Ancillary Response Files (\emph{arf}) and Redistribution Matrix Files (\emph{rmf}) \citep{Campana_2020} has been built to take this effect into account, and was used to evaluate the average count rate in the 3--11.5~keV energy range for the Crab Nebula+Pulsar ($C_{\mathrm{Crab}}$)
and the CXB ($C_{\mathrm{BKG}}$).

\subsection{Analytical determination of the Crab 
exposure time}

The dominant practical limitation is the on–axis exposure that can accumulate over time. In this section,, we present the analytical treatment for deriving a relation linking the expected $\chi^{2}$ of the Crab pulse profile to the observing time $T_\mathrm{exp}$.

\begin{figure}[h!]
    \centering
    \includegraphics[width=0.6\linewidth]{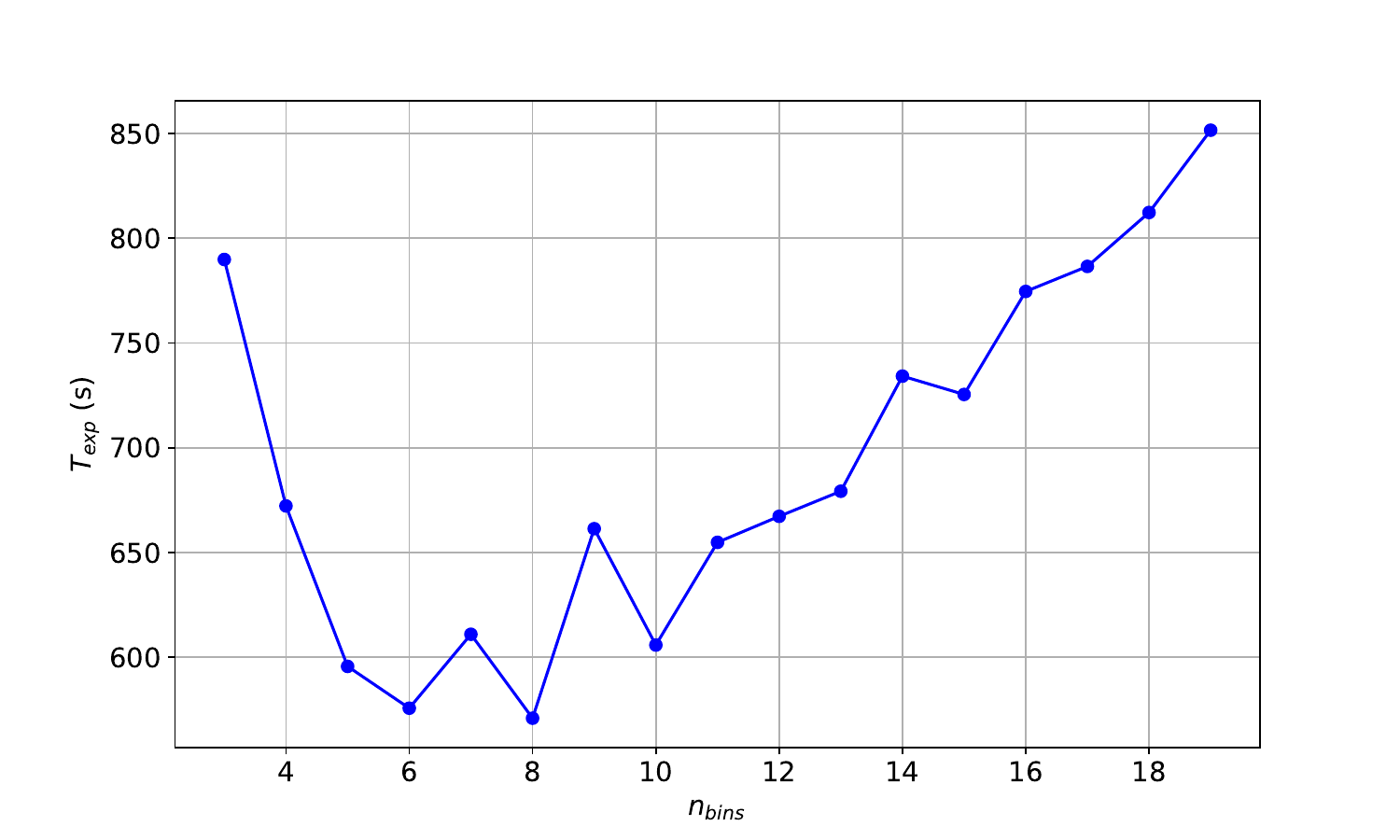}
    \caption{Required exposure time to obtain a $\chi^2$ threshold for a $3\sigma$ level detection as function of $n_\mathrm{bins}$. Although analytically derived, the observed scatter in \autoref{fig:3sigma_significance_level} arises because the solution is applied to an observed profile with finite counts and is therefore subject to statistical (Poisson) fluctuations.}
    \label{fig:3sigma_significance_level}
\end{figure}

\autoref{appendix:Analytical} shows the analytical solution to the problem, and by solving the \autoref{eq:X^2_analytical} for $T_\mathrm{exp}$, we find that the $\chi^2$ value can be expressed as:

\begin{equation}
    \chi^2_\mathrm{th} = T_\mathrm{exp} \cdot C_\mathrm{Crab}
    \cdot \sum_{i=1}^{n_\mathrm{bins}} \frac{(\text{FRAC}_i - \Delta \phi)^2}{\text{FRAC}_i + \Delta \phi \cdot \left( \frac{C_\mathrm{BKG}}{C_\mathrm{Crab}} \right)},
    \label{eq:X_tresh}
\end{equation}

where ${FRAC}_i$ is the per–bin pulsed fraction that maximizes the pulse significance, as defined in \autoref{appendix:Pulsation significance maximization} and \autoref{eq:FRAC_i} and $\Delta \phi$ the dimension of the phase bin.

We now assume, as null hypotheses, a flat signal in our data ($H_0$ corresponds to no Crab pulsation detection). The probability that, for a given threshold $\chi^2_\mathrm{th,dof}$, the observed $\chi^2$ is greater, comes from a well-known function $\rm P(\chi^2\geq\chi^2_\mathrm{th,dof})$, for a certain number of degrees of freedom dof.

\autoref{eq:X_tresh} can be inverted to derive the $T_\mathrm{exp}$ necessary to get the desired $\chi^2_\mathrm{th,dof}$ for a given number of degrees of $n_\mathrm{bins}$. We compute the $\chi^2_\mathrm{th,dof}$ to reach a $3\sigma$ significance level for each bin, and obtain the trend shown in \autoref{fig:3sigma_significance_level}.


\subsection{The Monte Carlo validation}
\label{appendix:Chi_distribution}

The previous discussion is associated with the $\chi^2$ threshold and so with the probability $\rm P(\chi^2\geq\chi^2_\mathrm{th,dof})$. This establishes the required exposure time to reach a minimum significance limit.

In order to verify the accuracy of the previous solution and the expected $\chi^2$ distribution as a function of a given $T_\mathrm{exp}$, we perform a simulation-based analysis starting from a theoretical profile with $n_\mathrm{bins}=15$ as shown in \autoref{fig:Shifted_pulse_profile}.

We adopt a fixed pulse shape template and denote its amplitude by \(I_p\) (in counts) as the sum of the CRAB and background contributions:

\begin{equation}
   I_{p} = I_\mathrm{ratio} \cdot T_\mathrm{exp} \cdot C_\mathrm{Crab} + \frac{1}{n_\mathrm{bins}} \cdot
   T_\mathrm{exp} \cdot C_\mathrm{BKG},
    \label{eq:pulse_profile}
\end{equation}

where $I_\mathrm{ratio}$ is the ratio of the counts in every bin of the RXTE profile over the total counts as shown in \autoref{fig:Shifted_pulse_profile}. When using the \autoref{eq:pulse_profile}, we extract the value of $I_{p}$ with a Poisson distribution, in order to achieve a series of Poisson realization of the Crab pulse profile at a given exposure time.

According to the observation exposure, in \autoref{eq:pulse_profile}, we adopt $T_\mathrm{exp}=730$~s and for each Poisson realization we evaluate the $\chi^2$ with respect to the average of the pulse profile. 

\begin{figure}[h!]
    \centering
    \includegraphics[width=0.6\linewidth]{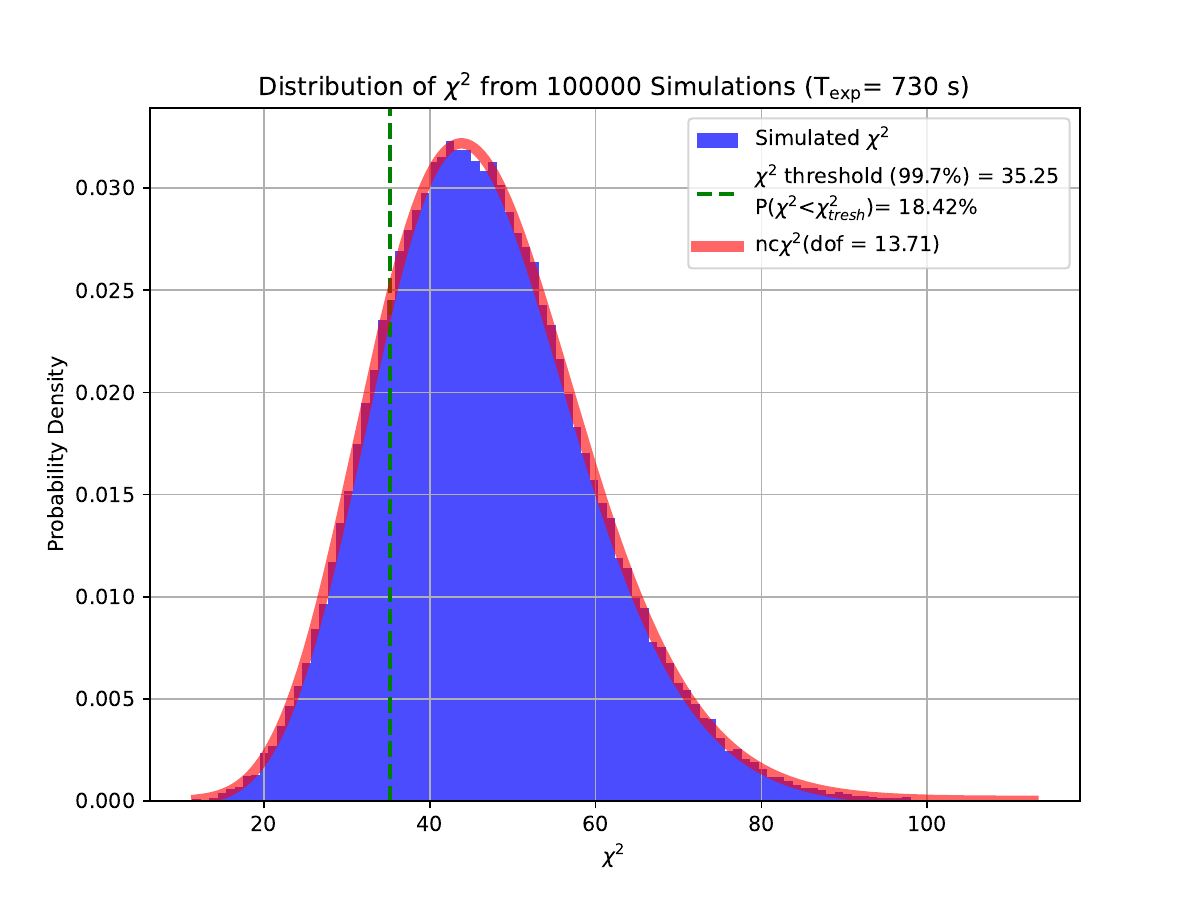}
    \caption{Monte Carlo distribution of the $\chi^{2}$ statistic for simulated Crab-like pulse profiles with $n_{\rm bins}=15$. The histogram is fitted with the probability density in \autoref{eq:ncX}. The vertical green line marks the $3\sigma$ detection threshold: for ${\rm dof}=14$ we adopt the threshold $\chi^{2}=35.25$, i.e. the 99.7\% upper-tail critical value of the central $\chi^{2}$ distribution.}
    \label{fig:Chi_distribution}
\end{figure}
\section{The analytical solution}
\label{appendix:Analytical}

The Monte Carlo distribution is shown in \autoref{fig:Chi_distribution} is fitted with a non-central $\chi^2$ (nc$\chi^2$) distribution:

\begin{equation}
f(x, dof, \lambda) = \frac{1}{2} \exp\left(-\frac{\lambda + x}{2}\right) \left(\frac{x}{\lambda}\right)^{(dof - 2)/4} L_{(dof - 2)/2} \left( \sqrt{\lambda x} \right),
\label{eq:ncX}
\end{equation}

where $\lambda$ is the non-centrality parameter. $L_{(k - 2)/2}$ denotes the modified Bessel function of first order of degree and $x$ the $\chi^2$. The parameters from the nc$\chi^2$ fit match expectations, and the dof are in line with the theoretical value $n_\mathrm{bins}-1=15-1=15$.

Therefore, we expect that in $\approx81.58 \%$ of cases, a 730~s on-axis \emph{HERMES} observation grants at least a 3$\sigma$ detection. 

The total detected count rate (in counts/s) is given by the sum of the Crab emission (nebula + pulsar) $C_\mathrm{Crab}$ and the background in the \emph{HERMES} FoV $C_\mathrm{BKG}$:
\begin{equation}
    C_\mathrm{tot} = C_\mathrm{Crab} + C_\mathrm{BKG}
\end{equation}

\begin{figure}[h!]
    \centering
    \includegraphics[width=0.7\linewidth]{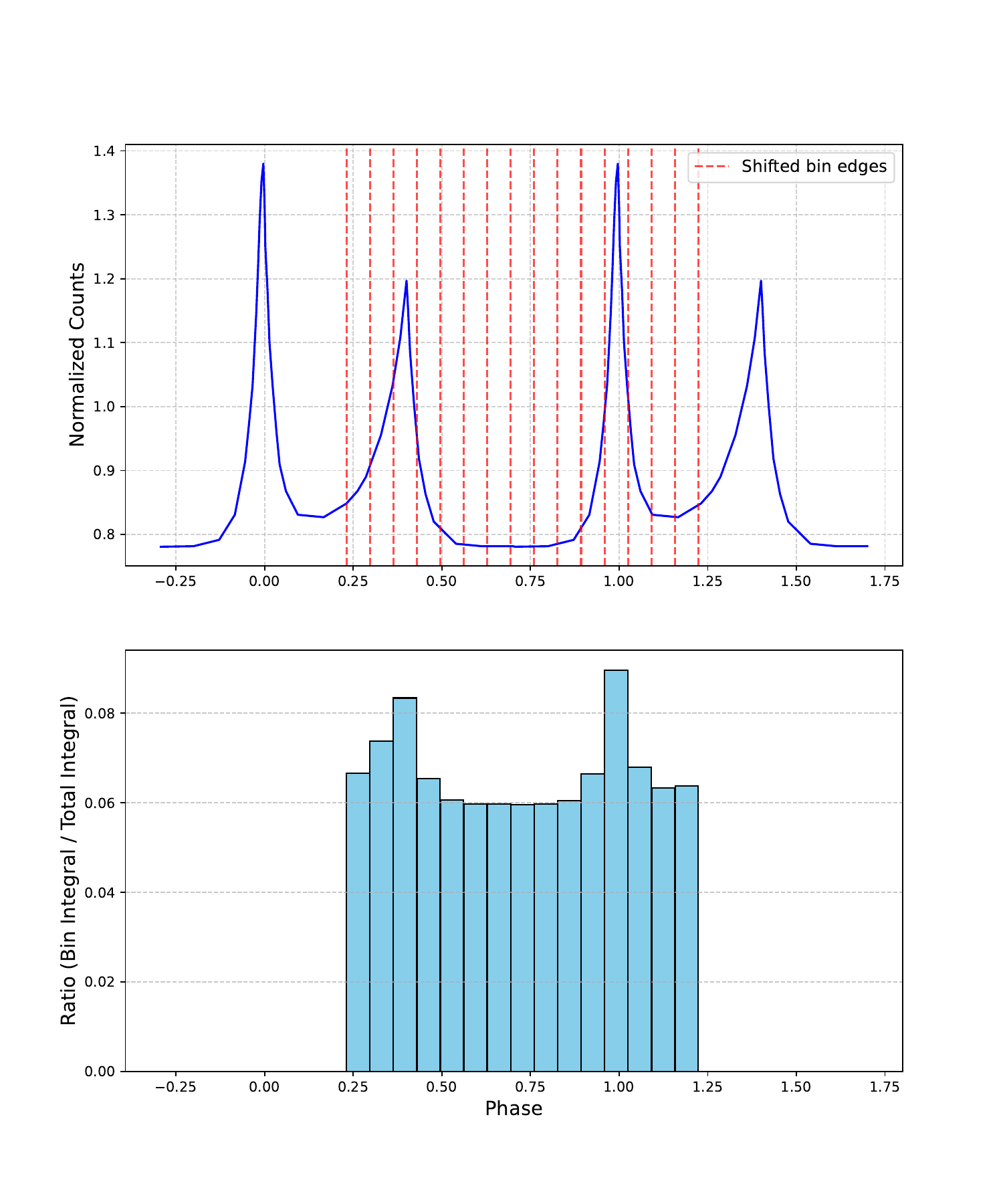}
    \caption{Upper plot: RXTE Crab Pulsar pulse profile is rebinned to obtain a 15-channel profile, for the given best shift. Lower plot: Normalized count ratio as evaluated in the red bins, with respect to total counts of the profile ($FRAC_i$). In this case, the on-pulse peaks are the 3rd and the 12th, the others are considered as off-pulse bins.}
    \label{fig:Shifted_pulse_profile}
\end{figure}

If we divide the pulse profile into $n_\mathrm{bins}$ bins, as in \autoref{fig:Shifted_pulse_profile}, the total rate in each phase bin $\phi_i$ is:
\begin{equation}
    C_\mathrm{tot}(\phi_i) = 
    \Delta\phi \cdot C_\mathrm{BKG} + \text{FRAC}_i \cdot C_{\text{CRAB}} \quad \text{(counts/s)}, 
\end{equation}
where $\Delta\phi$ is the bin width and $\text{FRAC}_i$ the count ratio evaluated for each bin as in \autoref{fig:Shifted_pulse_profile}. 

For an observation time $T_\mathrm{exp}$, the expected counts in each phase bin are:
\begin{equation}
    N_\mathrm{tot}(\phi_i) = C_\mathrm{tot}(\phi_i) \cdot T_\mathrm{exp},
\end{equation}

with an associated Poissonian error of:
\begin{equation}
\sigma_{N_i} = \sqrt{N_\mathrm{tot}(\phi_i)}
\end{equation}

The average total value of the expected counts is computed as:
\begin{equation}
    \overline{N_\mathrm{tot}} = \sum_{i=1}^{n_\mathrm{bins}} \frac{N_\mathrm{tot}(\phi_i)}{n_{\text{bins}}} \times \Delta\phi,
\end{equation}

where $\Delta\phi_{BINS}$ is the peak dimension in unit of phase.
We can now compute the theoretical chi-square value as:
\begin{equation}
    \chi^2_\mathrm{th} = \sum_{i=1}^{n_\mathrm{bins}} \frac{\left(N_\mathrm{tot}(\phi_i) - \overline{N_\mathrm{tot}}\right)^2}{N_\mathrm{tot}(\phi_i)}.
    \label{eq:X^2_analytical}
\end{equation}

\section{Pulsation significance maximization}
\label{appendix:Pulsation significance maximization}

To maximize the pulsation significance, we slide the edges of phase bins across the folded profile as in \autoref{fig:Shifted_pulse_profile}.  

For every trial position we compute, in each energy channel, the ratio

\begin{equation}
    \mathrm{FRAC}_i \;=\; \frac{N_i}{N_{\text{on}} + N_{\text{off}}}, 
\qquad i = 1,\dots, n_\mathrm{bins},
\label{eq:FRAC_i}
\end{equation}

i.e. the ratio of counts falling in each bin window  \(N_{i}\) to the total counts on–pulse (\(N_{\text{on}}\)) plus off–pulse (\(N_{\text{off}}\)), as defined in \autoref{fig:Shifted_pulse_profile}. 

The optimal phase shift is found by maximizing the variance

\begin{equation}
\sigma^{2} \;=\frac{1}{n_\mathrm{bins}}\; \sum_{i=1}^{n_\mathrm{bins}} \bigl( 1 - \mathrm{FRAC}_i \bigr)^{2},
\end{equation}

which reaches its maximum when the binning is such that two bins are centered on the main emission peaks of the pulse profile.  

Applying this optimization for different numbers of channels, in the 3--11.5~keV energy band, we produce a lookup table that records, the phase boundaries and the corresponding \(\mathrm{FRAC}_i\) for each profile represented with a given bins number.  

\end{document}